\documentclass[draft]{agujournal2019}
\usepackage{url} 
\usepackage{lineno}
\usepackage[finalnew]{trackchanges} 
\usepackage{soul}
\usepackage{natbib}
\usepackage{multirow}	
\usepackage[normalem]{ulem} 

\draftfalse
\journalname{JGR: Atmospheres}

\begin{document}

\title{A new full 3D model of cosmogenic tritium $^3$H production in the atmosphere (CRAC:3H)}

\authors{S.~V. Poluianov\affil{1,2}, G.~A. Kovaltsov\affil{3}, I.~G. Usoskin\affil{1,2}}

\affiliation{1}{Sodankyl\"a Geophysical Observatory, University of Oulu, Finland}
\affiliation{2}{Space Physics and Astronomy Research Unit, University of Oulu, Finland}
\affiliation{3}{Ioffe Physical-Technical Institute, St. Petersburg, Russia}

\correspondingauthor{Stepan Poluianov}{stepan.poluianov@oulu.fi}

\begin{keypoints}
\item A new CRAC:3H model of cosmogenic tritium ($^3$H) production in the atmosphere is presented.
\item For the first time, it provides 3D production, also explicitly treating particles heavier than protons.
\item This model provides a useful tool for the use of $^3$H as a tracer of atmospheric and hydrological circulation.
\end{keypoints}

\begin{abstract}
A new model of cosmogenic tritium ($^3$H) production in the atmosphere is presented.
The model belongs to the CRAC (Cosmic-Ray Atmospheric Cascade) family and is named as CRAC:3H.
It is based on a full Monte-Carlo simulation of the cosmic-ray induced atmospheric cascade using the Geant4 toolkit.
The CRAC:3H model is able, for the first time, to compute tritium production at any location
 and time, for any given energy spectrum of the primary incident cosmic ray particles, explicitly treating,
 also for the first time, particles heavier than protons.
This model provides a useful tool for the use of $^3$H as a tracer of atmospheric and hydrological circulation.
A numerical recipe for practical use of the model is appended.
\end{abstract}


\section{Introduction}

Tritium ($^3$H\remove{, earlier also called \textit{triton}}) is a radioactive isotope of hydrogen with the half-life time of
 approximately 12.3 years.
As an isotope of hydrogen, it is involved in the global water cycle and forms a very useful tracer of atmospheric
 moisture \citep[e.g.,][]{froehlich10,juhlke20} or hydrological cycles \citep{michel05}.
In the natural environment, tritium is mostly produced by galactic cosmic rays (GCR) in the atmosphere, as a sub-product of the induced
 nucleonic cascade, and is thus a cosmogenic radionuclide.
On the other hand, tritium is also produced artificially in thermonuclear bomb tests.
Before the nuclear-test ban became in force, a huge amount of tritium had been produced
 artificially and realised into the atmosphere, leading to an increase of the global reservoir inventory of tritium by two orders
 of magnitude above the natural level \citep[e.g.,][]{froehlich10,cauguoin16}.
Thus, the cosmogenic production of tritium was typically neglected as being too small against anthropogenic one.
However, as nearly 60 years have passed since the nuclear tests, its global content
 has reduced to the natural pre-bomb level \citep{palcsu18} and presently is mostly defined by
 the cosmogenic production.
Accordingly, natural variability of the isotope production can be again used \change{}{in
atmospheric tracing, water vapour transport, dynamics of the stratosphere-troposphere
exchanges over Antarctica} \citep{cauquoin15,fourre18,palcsu18,juhlke20,laszlo20}.
\change{}{Moreover, a combination of the $^3$H data with other tracers like atmospheric $^{10}$Be, 
which is also produced by cosmic-ray spallation reactions, 
but whose transport is different, can be a very powerful research tool.}
For this purpose, a reliable production model is needed, which is able to provide a full 3D and time
 variable production of tritium in the atmosphere.

Some models of tritium production by cosmic rays (CR) in the atmosphere \change{}{were} developed earlier.
First models \citep{fireman53,craig61,Nir_RGSP1966,lal67,OBrien_JGR1979} were based on simplified numerical or
 semi-empirical methods of modelling the cosmic-ray induced atmospheric cascade.
Later, a full Monte-Carlo simulation of the cosmogenic isotope production in the atmospheric cascade had been
 developed \citep{Masarik_JGR1999} leading to higher accuracy of the results.
However, that model had some significant limitations: (1) were considered only GCR protons
 (heavier GCR species were treated as scaled protons); (2) the energy spectrum of GCR was prescribed;
 (3) only global and latitudinal zonal mean productions were presented, implying no spatial resolution.
That model was slightly revisited by \citet{Masarik_JGR2009}, but the methodological
 approach remained the same.
A more recent tritium production model developed by \citet{Webber_JGR2007} is also based on a full
 Monte-Carlo simulation of the atmospheric cascade and was built upon the yield-function approach which
 allows dealing with any kind of the cosmic-ray spectrum.
However, only columnar (for the entire atmospheric column) production was provided by those authors,
 making it impossible to model the height distribution of isotope production.
Moreover, that model was dealing with CR protons only, while the contribution of heavier species
 to cosmogenic isotope production can be as large as 40\% (see section~\ref{s:prod}).

Here we present a new model of cosmogenic tritium production in the atmosphere, that is based on
 a full simulation of the cosmic-ray induced atmospheric cascade.
This model belongs to the CRAC (Cosmic-Ray Atmospheric Cascade) family and is named as \textit{CRAC:3H}.
The CRAC:3H model is able, for the first time, to compute tritium production at any location
 and time, for any given energy spectrum of the primary incident CR particles, explicitly treating,
 also for the first time, particles heavier than protons.
This model provides a useful tool for the use of $^3$H as a tracer of atmospheric and hydrological circulation.

\section{Production model}

The local production rate $q$ of a cosmogenic isotope, in atoms per second per gram of air,
 at a given location with the geomagnetic rigidity cutoff $P_{\rm c}$ and the atmospheric depth $h$ can be 
written as
\begin{equation}
q(h, P_{\rm c}) = \sum_{i}{ \int_{E_{{\rm c},{i}}}^{\infty} J_i(E)\cdot Y_i(E,h)\cdot dE},
\label{eq:q}
\end{equation}
where $J_i(E)$ is the intensity of incident cosmic-ray particles of the $i$-th type (characterized by the charge $Z_i$
 and atomic mass $A_i$ numbers) in units of particles per (s sr cm$^2$ GeV),
 $Y_i(E,h)$ is the isotope yield function in units of (atoms sr cm$^2$ g$^{-1}$, --- see section \ref{s:yf} for details),
 $E$ is the kinetic energy of the incident particle in GeV, $h$ is the atmospheric depth in units of (g/cm$^{2}$),
 $E_{c,i}=\sqrt{\left ({Z_i}\cdot P_{\rm c}/A_i\right )^2+E_0^2}-E_0$ is the energy corresponding to the local
 geomagnetic cutoff rigidity for a particle of type $i$, and the summation is over the particle types.
 $E_0=0.938$ GeV is the proton's rest mass.
The geomagnetic rigidity cutoff $P_{\rm c}$ quantifies the shielding effect of the geomagnetic field and can be
 roughly interpreted as a rigidity/energy threshold of primary incident charged particles required to
 imping on the atmosphere \citep[see formalism in][]{Elsasser_N1956,shea00}.

\subsection{Production function}
\label{s:yf}

Here we computed the tritium production function in a way similar to our previous works in the framework of
 the CRAC-family models \citep[e.g.,][]{Usoskin_JGR2008, Kovaltsov_EPSL2010, Kovaltsov_EPSL2012, Poluianov_JGR2016},
  viz. by applying a full Monte-Carlo simulation
 of the cosmic-ray induced atmospheric cascade, as briefly described below.
Full description of the nomenclature and numerical approach is available in \citet{Poluianov_JGR2016}.

The yield function $Y_i(E,h)$ (see equation~\ref{eq:q}) of a nuclide of interest provides the number of
 atoms produced in the unit (1 g/cm$^2$) atmospheric layer by incident particles of type $i$ (e.g., cosmic
 ray protons, $\alpha$-particles, heavier species) with the fixed energy $E$ and the unit intensity
  (1 particle per cm$^2$ per steradian).
The yield function should not be confused with the so-called production function $S_i(E,h)$,
 which is defined as the number of nuclide atoms produced in the unit atmospheric layer per one incident particle
 with the energy $E$.
In a case of the isotropic particle distribution, these quantities are related as
\begin{equation}
Y = \pi S,
\end{equation}
where $\pi$ is the conversion factor between the particle intensity in space and the particle flux
 at the top of the atmosphere \citep[see, e.g., chapter 1.6.2 in][]{Grieder2001}.

\begin{figure}
\center
\noindent\includegraphics[width=1\textwidth]{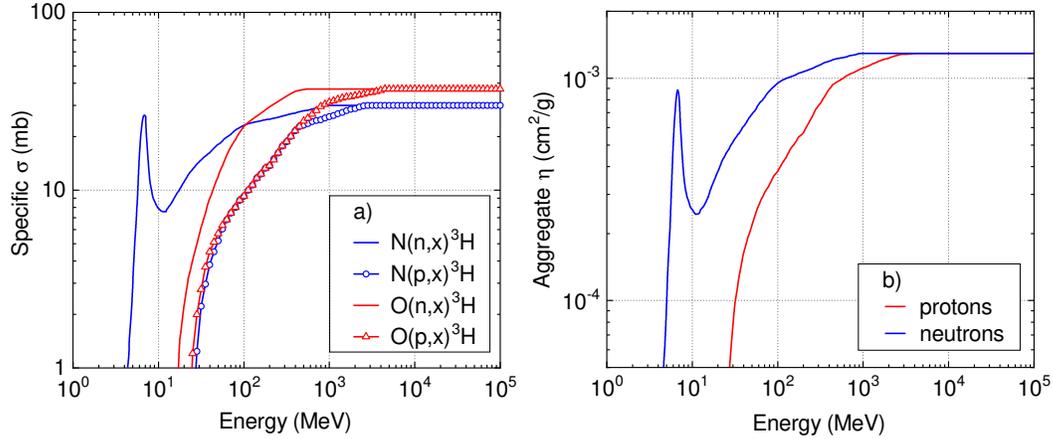}
\caption{Specific $\sigma$ (panel a) adopted from \citet{Nir_RGSP1966,Coste_AnA2012} and \textit{aggregate}
 $\eta(E)$ (panel b) cross-sections for production of tritium as a function of the particle's energy.}
\label{f:cs}
\end{figure}

The production function in units (atoms cm$^{2}$/g) can be calculated, for the isotropic flux of primary CR particles of type $i$, as
\begin{equation}
S_i(E,h) = \sum_l \int_0^E \eta_l(E') \cdot N_{i,l}(E,E',h)\cdot v_l(E') \cdot dE',
\label{eq:yf}
\end{equation}
where summation is over types $l$ of secondary particles of the cascade (can be protons, neutrons, $\alpha$-particles),
 $\eta_l$ is the `\textit{aggregate}' cross-section (see below) in units (cm$^2$/g),
 $N_{i,l}(E,E',h)$ and $v_l(E')$ are concentration and velocity of the secondary particles of type $l$ with energy $E'$
 at depth $h$.
The aggregate cross-section $\eta_l(E')$ is defined as
\begin{equation}
\eta_l(E') = \sum_j \kappa_j \cdot \sigma_{j,l}(E'),
\label{eq:eta}
\end{equation}
where $j$ indicates the type of a target nucleus in the air (nitrogen and oxygen for tritium),
 $\kappa_j$ is the number of the target nuclei of type $j$ in one gram of air,
 $\sigma_{j,l}(E')$ is the total cross-section of nuclear reactions
 between the $l$-th atmospheric cascade particle and the $j$-th target nucleus
 yielding the nuclide of interest.
Atmospheric tritium is produced by spallation of target nuclei of nitrogen and oxygen, which have the
 values of $\kappa_{\rm N} = 3.22\cdot10^{22}$ g$^{-1}$ and $\kappa_{\rm O} = 8.67\cdot10^{21}$ g$^{-1}$, respectively.
The reactions yielding tritium are caused mainly by the cascade neutrons and protons and include:
N(n,x)$^3$H;
N(p,x)$^3$H;
O(n,x)$^3$H;
O(p,x)$^3$H.
The cross-sections used here were adopted from \citet{Nir_RGSP1966} and \citet{Coste_AnA2012},
 as shown in Figure \ref{f:cs}a.
We assumed that cross-sections of the neutron-induced reactions are similar to those for protons above the energy of 2 GeV.
For reactions caused by $\alpha$-particles, N($\alpha$,x)$^3$H and O($\alpha$,x)$^3$H, the cross-sections were assessed from
 proton ones according to \citet{Tatischeff_APJS2006}.
These reactions are induced mostly by $\alpha$-particles from the primary CRs and are, hence, important only in the
 upper atmospheric layers.

The tritium aggregate cross-sections $\eta$ (equation~\ref{eq:eta}) are shown in Figure \ref{f:cs}b.
Although production efficiencies of protons and neutrons are similar at high energies, they differ significantly
 in the $<$500 MeV range.
Because of the lower energy threshold and higher cross-sections for neutrons in this energy range, comparing to protons,
 tritium production is dominated by neutrons in a region where the cascade is fully-developed,
 viz., in the lower part of the atmosphere.

The quantity $N_{i,l}(E,E',h)\cdot v_l(E')$ describing 
the cascade particles (equation~\ref{eq:yf}) was computed using
 a full Monte Carlo simulation of the cascade induced in the atmosphere by energetic cosmic-ray particles.
The general computation scheme was similar to that applied by \citet{Poluianov_JGR2016}.
The simulation code was based on the Geant4 toolkit v.10.0 \citep{Geant4-1, Geant4-2}.
In particular, we used the physics list QGSP\_BIC\_HP (Quark-Gluon String model for high-energy interactions;
 Geant4 Binary Cascade model; High-Precision neutron package) \citep{Geant4-Phys2013}, which was shown to
 describe the cosmic ray cascade with sufficient accuracy \citep[e.g.,][]{mesick18}.
We simulated a real-scale spherical atmosphere with the inner radius of 6371 km,
 height of 100 km and thickness of 1050 g/cm$^2$.
The atmosphere was divided into homogeneous spherical layers with the thickness ranging from 1 g/cm$^2$ (at the top) to
 10 g/cm$^2$ near the ground.
The atmospheric composition and density profiles were taken according to the atmospheric model NRLMSISE-00 \citep{Picone2002}.
Cosmic rays were simulated as isotropic fluxes of mono-energetic protons and $\alpha$-particles, while heavier
 species were considered as scaled $\alpha-$particles (see section~\ref{s:GCR}).
The simulations were performed with a logarithmic grid of energies between 20 MeV/nuc and 100 GeV/nuc.
The number of simulated incident particles was set so that the statistical accuracy of the isotope production
 should be better than 1\% in any location.
This number varied from 1000 incident particles for $\alpha-$particles with the energy of 100 GeV/nucleon
 to $2\cdot 10^7$ simulations for 20-MeV protons.
The results were extrapolated to higher energies, up to 1000 GeV/nuc, by applying a power law.
The yield of the secondary particles (protons, neutrons and $\alpha$-particles) at the top of each atmospheric
 layer was recorded as histograms with the spectral (logarithmic) resolution of 20 bins per one order of
 magnitude in the range of the secondary particle's energy between 1 keV and 100 GeV.
The primary CR particles were also recorded in the same way.

\begin{figure}
\center
\noindent\includegraphics[width=1\textwidth]{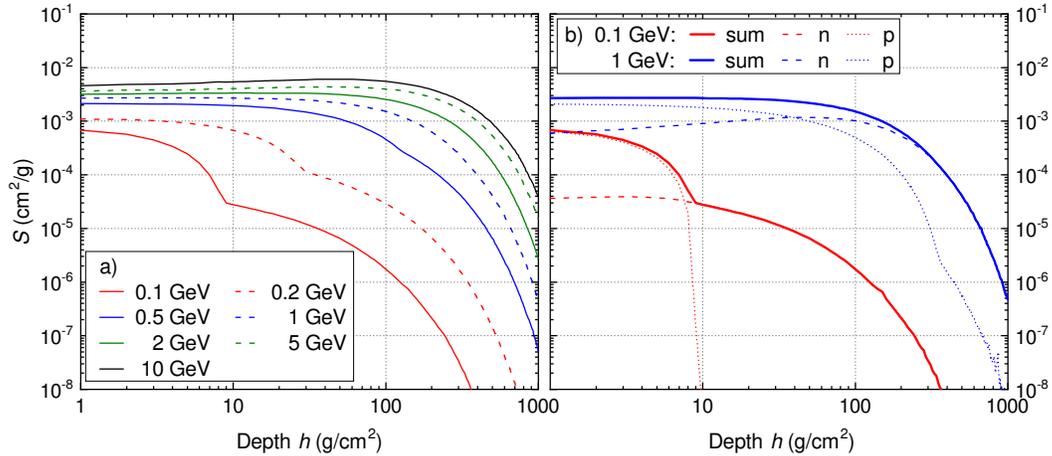}
\caption{Production function $S$=$Y/\pi$ of tritium by primary protons.
Panel a: production function $S$ by primary protons with energies between 0.1--10 GeV, as denoted in the legend.
Panel b: contribution of protons (p) and secondary neutrons (n) to the production function (sum)
 for 0.1 GeV (red) and 1 GeV (blue) primary protons.
}
\label{f:yf_depth}
\end{figure}

The production functions $S_i(E,h)$ were subsequently calculated from the simulation results, using equation (\ref{eq:yf}),
 for a prescribed grid of energies and atmospheric depths and are tabulated in the Supporting Information.
Some examples of the tritium production function are shown in Figure~\ref{f:yf_depth}
 for primary CR protons.
One can see in Figure \ref{f:yf_depth}a that the efficiency of tritium atom production grows with
 the energy of the incident particles because of larger atmospheric cascades induced.
Contributions of different components to the total production are shown in Figure~\ref{f:yf_depth}b for
 low (0.1 GeV) and medium (1 GeV) energies of the primary proton.
The red curve for the 0.1 GeV incident protons depicts a double-bump structure: the bump in the upper
 atmospheric layers ($h<$10 g/cm$^2$) is caused by spallation reactions caused mostly by the primary protons
 (as indicated by the red dotted curve), while the smooth curve at higher depths is due to secondary
 neutrons (red dashed curve).
Overall, production of tritium at depths greater than 10 g/cm$^2$ is very small for the low-energy primary protons.
On the other hand, higher-energy (1 GeV, blue curves in Figure~\ref{f:yf_depth}b) protons effectively
 form a cascade reaching the ground, where the contribution of secondary neutrons dominates below
 $\approx 50$ g/cm$^2$ depths.

\begin{figure}
\center
\noindent\includegraphics[width=0.8\textwidth]{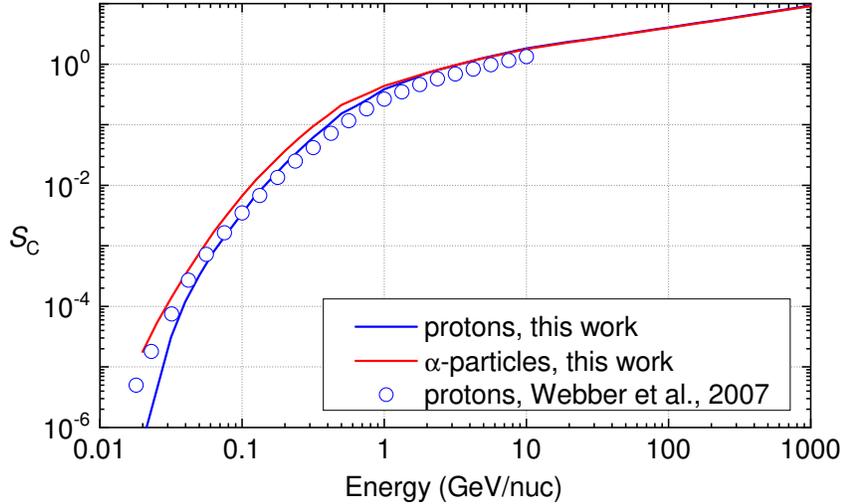}
\caption{Columnar production function $S_{\rm C}$=$Y_{\rm C}/\pi$ (number of atoms per primary incident nucleon)
 of tritium by incident protons (blue line) and $\alpha$-particles (red line).
 Tabulated values are available in \change{}{the Supporting Information}.
Circles indicate the production function for protons from \citet{Webber_JGR2007}.
}
\label{f:yf_column}
\end{figure}

This \change{}{type} of the depth/altitude profiles or the tritium production function was not studied in earlier works,
 where only columnar functions, viz. integrated over the full atmospheric column, were presented \citep{Webber_JGR2007}.
Therefore, in order to compare our results with the earlier published ones, we also calculated the columnar
 production function
\begin{equation}
S_{\rm C}(E) = \int_0^{h_{\rm sl}}{S(E,h)\cdot dh},
\end{equation}
where $h_{\rm sl}=1033$ g/cm$^2$ is the atmospheric depth at the mean sea level or the
 thickness of the entire atmospheric column.
The columnar production function is tabulated in \change{}{the Supporting Information} and 
depicted in Figure~\ref{f:yf_column}
 along with the earlier results published by \citet{Webber_JGR2007} for incident protons.
No results for incident $\alpha$-particles have been published earlier,
 and the production function of cosmogenic tritium by cosmic-ray $\alpha$-particles is
 presented here for the first time.
One can see that, while the production functions for incident protons generally agree between
 our work and the results by \citet{Webber_JGR2007}, there are some small but systematic differences.
In particular, our result is lower than that of \citet{Webber_JGR2007} in the low-energy range
 below 100 MeV.
It should be noted that the contribution of this energy region to the total production of tritium is
 negligible because of the geomagnetic shielding in such a way that low-energy incident
 particles can impinge on the atmosphere only in spatially small polar regions.
In the energy range above 200 MeV, the tritium production function computed here is higher
 than that from \citet{Webber_JGR2007}.
The difference is not \change{}{large}, $\approx 30$\%, but systematic and can be related to the
 uncertainties in the cross-sections or details of the cascade simulation (FLUKA vs. Geant4).
Overall, our model predicts slightly higher production of tritium than the one by \citet{Webber_JGR2007},
 for the same cosmic-ray flux.

\subsection{Cosmic-ray spectrum}
\label{s:GCR}
The first term $J_i(E)$ in equation (\ref{eq:q}) refers to the spectrum of differential intensity of the incident
 cosmic-ray particles.
A standard way to model the GCR spectrum for practical applications is based on the so-called
 \textit{force-field approximation} \citep{Gleeson_ApJ1967,Caballero-Lopez_JGR2004,Usoskin_phi_05},
 which parameterizes the spectrum with reasonable accuracy even during disturbed periods, as validated
 by direct in-space measurements \citep{usoskin_PAMELA_15}.
In this approximation, the differential energy spectrum of the $i$-th component of GCR near Earth
 (outside of the Earth's magnetosphere and atmosphere) is parameterized in the following form:
\begin{equation}
J_i(E,t) = J_{{\rm LIS},i}(E + \Phi_i(t))\frac{E(E + 2E_0)}{(E + \Phi_i(t))(E + \Phi_i(t) + 2E_0)},
\label{eq.FF}
\end{equation}
where $J_{{\rm LIS},i}$ is the differential intensity of GCR particles in the local interstellar medium,
 often called the local interstellar spectrum (LIS),
 $E$ is the particle's kinetic energy per nucleon, $E_0$ is the rest energy of a proton (0.938 GeV),
 and $\Phi_i(t)\equiv \phi(t) \cdot Z_i/A_i$ is the modulation parameter defined by the modulation potential $\phi(t)$
 as well as the charge ($Z_i$) and atomic ($A_i$) numbers of the particle of type $i$, respectively.
The spectrum at any moment of time $t$ is fully determined by a single time-variable parameter $\phi(t)$, which has the
 dimension of potential (typically given in MV or GV) and is called the modulation potential.
The absolute value of $\phi$ makes no physical sense and depends on the exact shape of LIS
 \citep[see discussion in][]{Usoskin_phi_05, Herbst_JGR2010, Herbst_JGR2017, Asvestari_JGR2017}.

In this work, we made use of a recent parameterization of the proton LIS \citep{Vos_ApJ2015}, which is partly based
 on direct \textit{in situ} measurements of GCR:
\begin{equation}
J_{{\rm LIS}}(E) = 0.27\,\frac{E^{1.12}}{\beta^2}
\left(\frac{E + 0.67}{1.67}\right)^{-3.93},
\label{eq:LIS}
\end{equation}
where $J_{{\rm LIS}}(E)$ is the differential intensity of GCR protons in the local interstellar medium
 in units of particles per (s sr cm$^2$ GeV), $E$ and $\beta=v/c$ are the particle's kinetic energy (in GeV)
 and the velocity-to-speed-of-light ratio, respectively.
Following a recent work \citep{Koldobskiy_JGR2019} based on a joint analysis of data from the space-borne experiment
 AMS-02 (Alpha Magnetic Spectrometer) and from the ground-based neutron-monitor network, we assumed
 that LIS (in the number of nucleons) of all heavier ($Z\geq$2) GCR species can be represented
 by the LIS for protons scaled with a factor of 0.353 for the same energy per nucleon.

The integral production rate in the entire atmospheric column is called the columnar production rate.
For a given location, characterized by the geomagnetic cutoff rigidity $P_{\rm c}$, and at the time moment $t$ 
 it is defined as
\begin{equation}
Q_{\rm C}(P_{\rm c},t) = \int_{0}^{h_{\rm sl}}q(h,P_{\rm c},t)\cdot dh.
\label{eq:int_q}
\end{equation}
The global production rate $Q_{\rm global}$ is the spatial average of $Q_{\rm C}(P_{\rm c})$ over the globe,
 while the integral of $Q$ over the globe yields the total production of tritium.

Production of tritium by GCR, which always bombard the Earth's atmosphere, is described above.
Production by solar energetic particles (SEP) can be computed in a similar way, with the SEP
 energy spectrum entering directly in equation (\ref{eq:q}).


\section{Results}
\label{s:prod}

Using the production function computed here (section~\ref{s:yf}) and applying 
equations~(\ref{eq:q}) and (\ref{eq:int_q}), we calculated the mean production rate $Q$ of tritium
  in the atmosphere for different levels of solar modulation (low, moderate and high),
  for the entire atmosphere and only for the troposphere.
The results are shown in Table \ref{t:prod_rate}.
The modeled local production rates $q(h,P_c)$ (equation \ref{eq:q}) used for the computation can be found
in a tabular form in the supporting information.
\begin{table}[t]
\center
\caption{Tritium production rates (in atoms/(s cm$^2$)) averaged globally (see also Figure~\ref{f:prod_curve})
 and over the polar regions (geographical latitude 60$^\circ$--90$^\circ$), separately in the entire atmosphere
 and only the troposphere for different levels of solar activity: low, medium and high ($\phi$=400, 650 and 1100 MV, respectively).
The values of the modulation potential correspond to the formalism described in section~\ref{s:GCR}.
The geomagnetic field is taken according to IGRF \citep[International Geomagnetic Reference Field,][]{Thebault_EPS2015}
 for the epoch 2015.
The tropopause height profile is adopted from \citet{Wilcox_QJRMS2012}.}
\begin{tabular}{l|cc|cc}
\hline
\multirow{2}{*}{Solar activity} & \multicolumn{2}{|c|}{Entire atm.} & \multicolumn{2}{|c}{Troposphere} \\
 & Global & Polar & Global & Polar  \\
\hline
Low    & 0.41 & 0.92 & 0.12 & 0.16 \\
Moderate & 0.345 & 0.72 & 0.11 & 0.14 \\
High  & 0.27 & 0.51 & 0.09 & 0.10 \\
\hline
\end{tabular}
\label{t:prod_rate}
\end{table}

The global production rate of tritium for \change{}{a moderate solar activity}
 ($\phi=650$ MV), which is the mean level for the modern epoch \citep{Usoskin_JGR2017}, is 0.345 atoms/(s cm$^2$).
This value can be compared with earlier estimates of the global production rate of tritium.
We performed a literature survey and found that the estimates performed before 1999 were based on different
 approximated approaches and vary by a factor of 2.5, between 0.14--0.36 atoms/(s cm$^2$)
 \citep{craig61,Nir_RGSP1966,OBrien_JGR1979,masarik95}.
Modern estimates, based on full Monte-Carlo simulations, are more constrained.
The early value of the global production rate of 0.28 atoms/(s cm$^2$) published by \citet{Masarik_JGR1999} was
 revised by the authors to 0.32 atoms/(s cm$^2$) in \citet{Masarik_JGR2009}.
Our value is very close to that, despite the different computational schemes and assumptions made.
The computed global production rate also agrees with the estimates obtained from reservoir inventories
 \citep[e.g.][]{craig61}, that are, however, loosely constrained within a factor of about four, between
 0.2--0.8 atoms/(s cm$^2$).
We note that heavier-than-proton primary incident particles contribute about 40\% to the global production of tritium,
 in the case of GCR, and thus, it is very important to consider these particles explicitly.

\begin{figure}
\center
\noindent\includegraphics[width=0.7\textwidth]{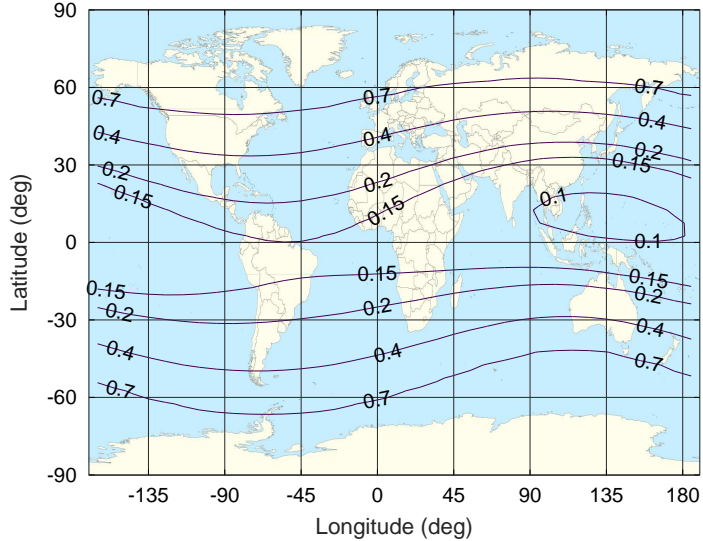} \vspace{0.5cm}
\caption{
Geographical distribution of the columnar production rate $Q_{\rm C}$ (atoms/(s cm$^2$)) of
 tritium by GCR corresponding to a moderate level of solar activity ($\phi$=650 MV).
The geomagnetic cutoff rigidities were calculated using the eccentric tilted dipole
 approximation \citep{Nevalainen2013} for the IGRF model (epoch 2015).
Other model parameters are as described above.
The background map is from Gringer/Wikimedia Commons/public domain.
}
\label{f:prod_map}
\end{figure}

Geographical distribution of the columnar production rate $Q_{\rm C}(P_{\rm c})$ of tritium is shown in Figure \ref{f:prod_map}.
It is defined primarily by the geomagnetic cutoff rigidity \citep[e.g.,][]{Smart2009, Nevalainen2013}
 and varies by an order of magnitude between the high-cutoff spot in the equatorial west-Pacific region and
 polar regions.

\begin{figure}[t]
\center
\noindent\includegraphics[width=0.8\textwidth]{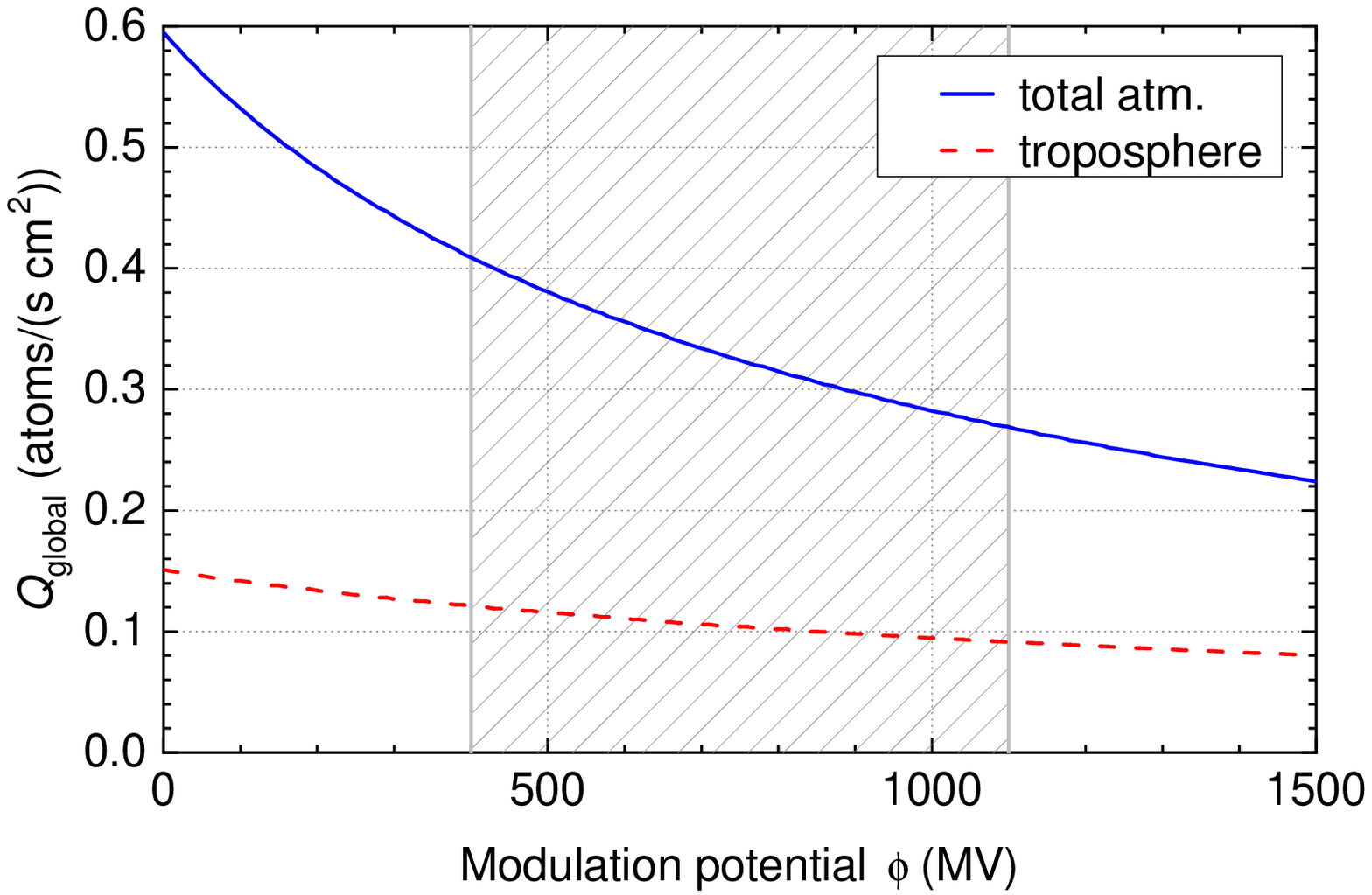}
\caption{Global columnar production $Q_{\rm global}$ of tritium, in the entire atmosphere and only in the
 troposphere, as a function of solar activity quantified via the heliospheric modulation potential.
The shaded area denotes the range of a solar cycle modulation for the modern epoch.
The geomagnetic field corresponds to the IGRF for the epoch 2015.
The tropopause height profile is adopted from \citet{Wilcox_QJRMS2012}.
The values of the modulation potential correspond to the formalism described in section~\ref{s:GCR}.
}
\label{f:prod_curve}
\end{figure}

Dependence of the global production rate of tritium on solar activity quantified via the
 modulation potential $\phi$ is shown in Figure~\ref{f:prod_curve}, both for the entire atmosphere
 and for the troposphere.
The tropospheric contribution to the global production is about 31\% on average,
 ranging from 30\% (solar minimum) to 34\% (solar maximum).

\begin{figure}
\center
\noindent\includegraphics[width=0.7\textwidth]{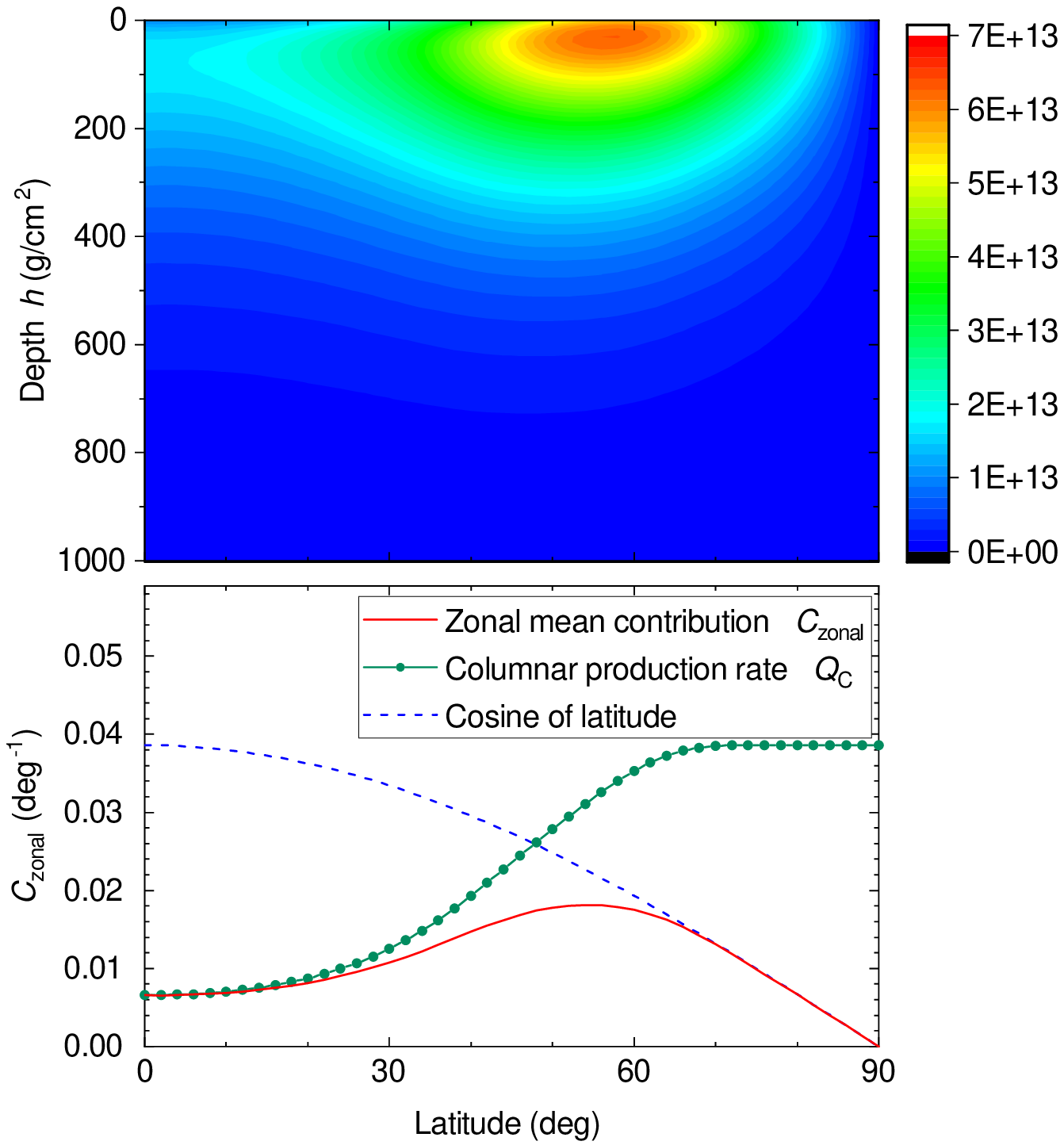}
\caption{
Upper panel: Tritium zonal production rate by GCR ($\phi$=650 MV, geomagnetic field IGRF epoch 2015) as 
\change{}{a} function of the atmospheric depth and northern geographical latitude.
The color scale (on the right) is given in units of atoms per second per degree of latitude per gram/cm$^2$.
Bottom panel: zonal mean contribution $C_{\rm zonal}$ (red curve, per degree of latitude) to the tritium global production rate
 (a columnar integral of the distribution shown in the upper panel), normalized so that its 
total integral over all latitudes is equal to unity.
Green dot-dashed and blue dashed lines represent the columnar production rate and cosine of latitude, respectively (both in
 arbitrary units), and $C_{\rm zonal}$ is directly proportional to their product.
}
\label{f:map_int}
\end{figure}

Even though the production rate is significantly higher in the polar region, its contribution to the
 global production is not dominant, because of the small area of the polar regions.
Figure~\ref{f:map_int} (upper panel) presents the production rate of tritium in latitudinal zones (integrated over longitude in one
 degree of geographical latitude) as a function of geographical latitude and atmospheric depth.
It has a broad maximum at mid-latitudes (40--70$^\circ$) in the stratosphere (10--100 g/cm$^2$ of depth)
 and ceases both towards the poles and ground.
The bottom panel of the Figure depicts the zonal mean contribution (red curve) of the entire atmospheric column
 into the total global production.
It illustrates that the distribution with a maximum at mid-latitudes shape is defined by two concurrent processes: 
 the enhanced production (green curve) and reduced zonal area (blue curve) from the equator to the pole.
The zonal contribution is proportional to the product of these two processes.

\begin{figure}
\center
\noindent\includegraphics[width=0.8\textwidth]{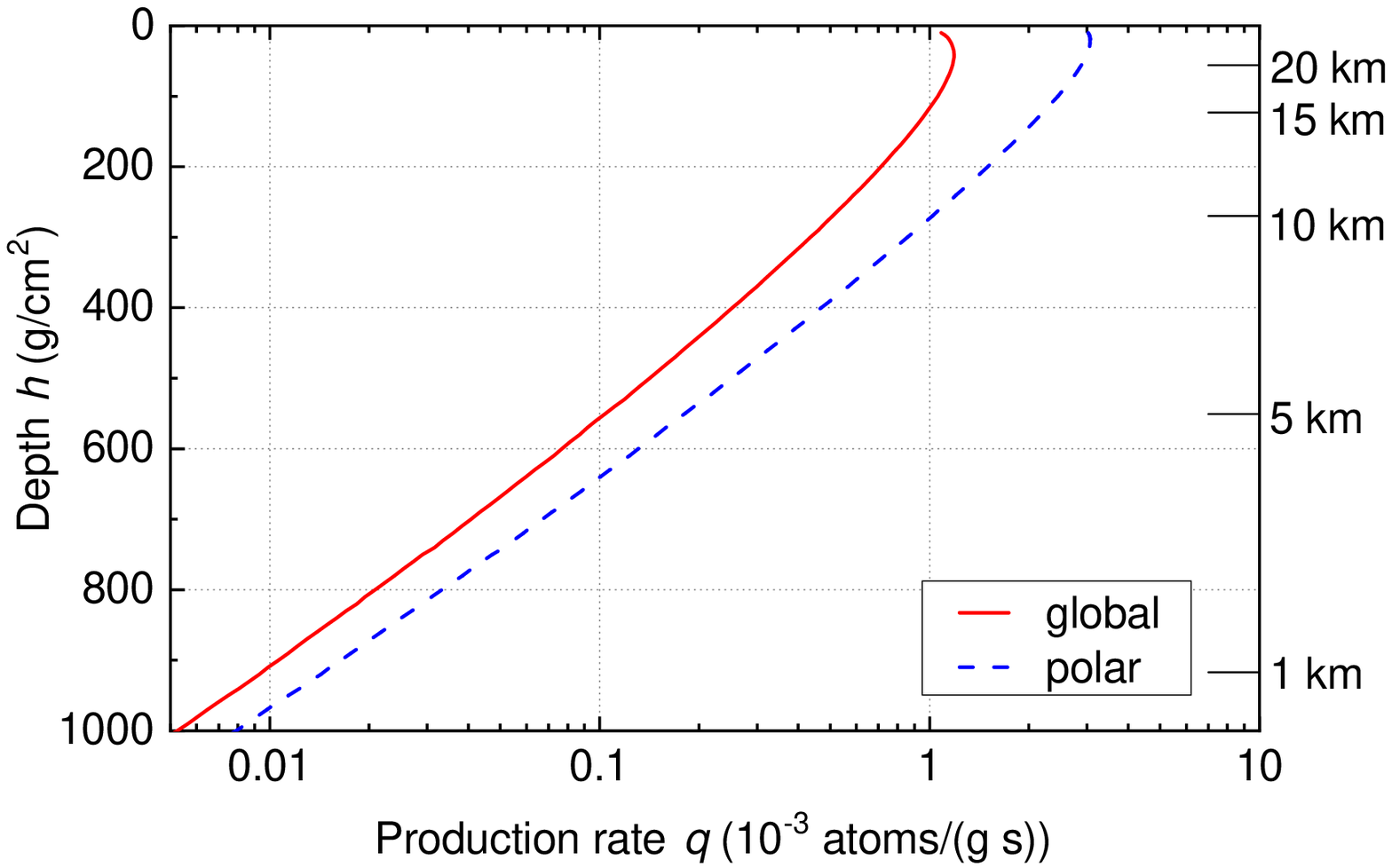}
\caption{
Altitude profile of the tritium differential production $q$ (equation~\ref{eq:q})
 by GCR for the moderate solar activity level ($\phi$=$650$ MV).
The red solid and blue dash lines represent the global and polar (60$^\circ$--90$^\circ$)
 production rates, respectively.
The horizontal marks on the right indicate the approximate altitude, which depends on the exact atmospheric conditions.
}
\label{f:prod_altprofile}
\end{figure}

The altitude profile of the tritium production rate by GCR for the moderate level of solar activity
 is shown in Figure \ref{f:prod_altprofile}.
The maximum of the globally averaged production is located at about 40 g/cm$^2$ or 20 km
 of altitude in the stratosphere, corresponding to the Regener-Pfotzer maximum where
 the atmospheric cascade is most developed.
The maximum of production is somewhat higher in the polar region because of the reduced
 geomagnetic shielding there, so that lower-energy CR particles can reach the location.

\begin{figure}
\center
\noindent\includegraphics[width=0.7\textwidth]{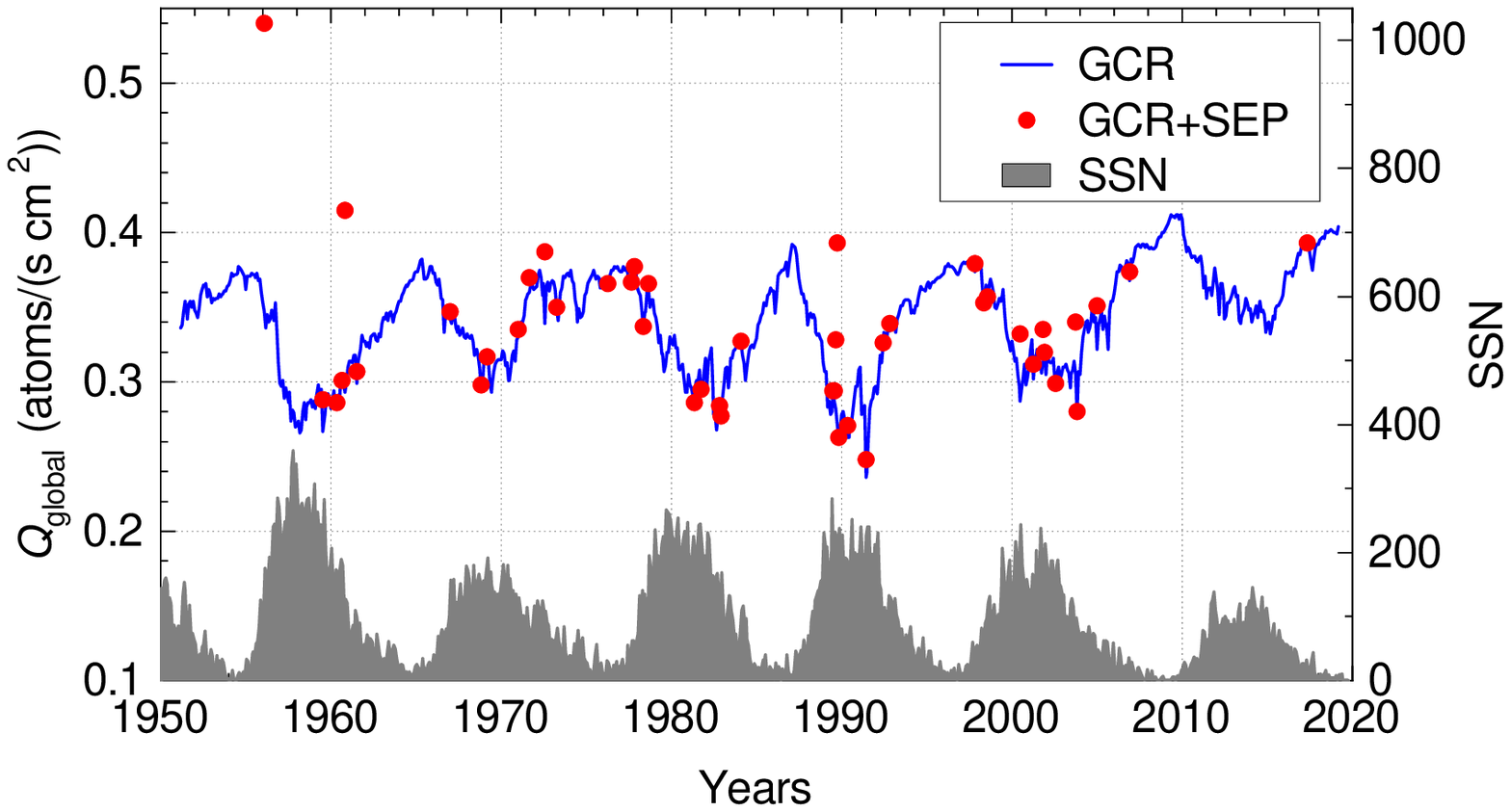}
\caption{Monthly means of the global production rates $Q_{\rm global}$ of tritium computed here for the period
 1951--2018.
The blue curve is for the GCR production (modulation potential and geomagnetic field were adopted from
 \citet{Usoskin_JGR2017} and IGRF, respectively).
The red dots indicate periods of GLE events (\url{http://gle.oulu.fi}) with additional production of tritium
 by SEPs as computed using the spectral parameters adopted from \citet{Raukunen_SWSC2018}.
The grey-shaded curve in the bottom represents the sunspot number (right-hand side axis) adopted from
 SILSO \cite[\url{http://www.sidc.be/silso/datafiles},][]{clette16}. }
\label{f:prod_SSN}
\end{figure}

Figure~\ref{f:prod_SSN} depicts temporal variability of the global tritium production for the
 period 1951--2018, computed using the model presented here.
To indicate the solar cycle shape, the sunspot numbers are also shown in the bottom.
The contribution from GCR is shown by the blue curve and computed using
 the modulation potential reconstructed from the neutron-monitor network \citep{Usoskin_JGR2017}.
Red dots consider also additional production of tritium by strong SEP events, identified as ground-level
 enhancement (GLE) events (\url{http://gle.oulu.fi}).
\change{}{This} is negligible on the long run but may contribute essentially
 on the short-time scale.
Overall, the production of tritium is mostly driven by the heliospheric modulation of GCR as
 implied by obvious anti-correlation with the sunspot numbers.




\section{Conclusion}
A new full model CRAC:3H of tritium cosmogenic production in the atmosphere is presented.
It is able to compute the tritium production rate at any location in 3D and for any type of the incident particle energy
 spectrum/intensity --- slowly variable galactic cosmic rays or intense sporadic events of solar energetic particles.
The core of the model is the yield/production function, rigorously computed by applying a full Monte-Carlo simulation
 of the cosmic-ray induced atmospheric cascade with high statistics and is tabulated in the Supporting Information.
Using this tabulated function, one can straightforwardly and easily calculate the production of tritium for any
 conditions in the Earth's atmosphere (see Appendix A), including solar modulation of GCR, sporadic SEP events, changes of the
 geomagnetic field, etc.
The columnar and global production of tritium, computed by the new model, is comparable with most recent estimates
 by other groups, but is significantly higher than the results of earlier models, published before 2000.
It also agrees well with empirical estimates of the tritium reservoir inventories, considering large uncertainties of the latter.
Thus, for the first time, a reliable model is developed \change{}{that} provides a full 
3D production of tritium in the atmosphere.
\change{}{These results can be used as an input for atmospheric transport models or for direct
comparison with tritium observations that are important for the study of solar activity in link with
the hydrological cycle or for evaluation of the atmospheric dynamics in models.}

\appendix
\section{Calculation of tritium production: Numerical algorithm}

Using the production function $S(E,h)$ presented here in the Supporting Information, one can
 easily compute tritium production at any given location (quantified by the local geomagnetic rigidity cutoff $P_{\rm c}$ and
 atmospheric depth $h$), and time $t$, following the numerical algorithm below.
\begin{enumerate}
\item
For a given moment of time $t$, the intensity of incident primary particles can be evaluated, in case of GCR,
 using equations (\ref{eq.FF}) and (\ref{eq:LIS}) for the independently known modulation potential $\phi$ (e.g., as
 provided at \url{http://cosmicrays.oulu.fi/phi/phi.html}).
These formulas can be directly applied for protons, while the contribution of heavier species ($Z\geq$2) can be
 considered, using the same formulas, but applying the scaling factor of 0.353 for LIS, which is given in
 number of nucleons, and considering kinetic energy per nucleon.
Thus, the input intensities of the incident protons $J_{\rm p}(E,t)$ and heavier species $J_{\rm \alpha}(E,t)$,
 the latter effectively including all heavier species, can be obtained.
Energy should be in units of GeV, and $J(E)$ in units of nucleons per (sr cm$^2$ s GeV).
The energy grid is recommended to be logarithmic (at least 10 points per order of magnitude).

\item
The production function $S_i(E,h)$ for the given atmospheric depth $h$ can be obtained, for both protons $S_{\rm p}$ and
 heavier species $S_\alpha$, from the Supporting Information in units of (cm$^2$/g).
The yield function is defined as $Y=\pi\cdot S$, in units of (sr cm$^2$/g), also separately for protons and
 heavier species.
The product of the yield function and the intensity of incident particles is called the response function
 $F_i(E,h)=Y_i(E,h)\cdot J_i(E)$, separately for protons and heavier species.

\item
As the next step, the local geomagnetic rigidity cutoff $P_{\rm c}$, which is related to the lower integration
 bound in equation~(\ref{eq:q}), needs to be calculated for a given location and time.
A good balance between simplicity and realism is provided by the eccentric tilted dipole approximation
 of the geomagnetic field \citep{Nevalainen2013}.
The value of $P_{\rm c}$ in this approximation can be computed using a detailed numerical recipe
 \cite[Appendix A in][]{Usoskin_JASTP2010}.
This approach works well with GCR, but is too rough for an analysis of SEP events, where a detailed
 computations of the geomagnetic shielding is needed \citep[e.g.,][]{Mishev2014}.

\item
Next, the response function $F_i$ should be integrated above the energy bound defined by the
 geomagnetic rigidity cutoff $P_{\rm c}$, as specified in equation (\ref{eq:q}) separately for the
 protons and $\alpha-$particles (the latter effectively includes also heavier $Z>$2 species).
Since the response function is very sharp, the use of standard methods of numerical integration, such as trapezoids,
 Gauss, etc., may lead to large uncertainties.
For numerical integration of equation (\ref{eq:q}), the piecewise power-law approximation is recommended,
 as described below.
Let function $F(E)$ whose values are defined at grid points $E_1$ and $E_2$ as $F_1$ and $F_2$, respectively, be approximated
 by a power law between these grid points.
Then its integral on the interval between these grid points is
\begin{equation}
\int_{E_1}^{E_2}{F(E)\cdot dE} = {(F_2\cdot E_2 - F_1\cdot E_1)\cdot \ln{(E_2/E_1)} \over \ln{(F_2/ F_1)} + \ln{(E_2/E_1)}}.
\end{equation}
The final production rate at the given location, atmospheric depth and time is the sum of the two components
 (protons and $\alpha-$particles).

\item
In a case when not only the very local production rate of tritium is required, but spatially integrated or averaged,
 the columnar production function (equation~\ref{eq:int_q}) can be used.
The spatially averaged/integrated production can be then obtained by averaging/integration over the appropriate area
 considering the changes in the geomagnetic cutoff rigidity $P_{\rm c}$.

\end{enumerate}


\acknowledgments
The yield/production functions of tritium, obtained in this work, are available in the
 Supporting Information to this paper. 
The used cross-section data can be found in 
\citet{Nir_RGSP1966} and \citet{Coste_AnA2012}.
The toolkit Geant4 \citep{Geant4-1, Geant4-2} is freely distributed under
Geant4 Software License at \url{http://www.geant4.org}.
This work used publicly available data for SEP events from the GLE database
(\url{http://gle.oulu.fi}), sunspot number series from SILSO 
\cite[\url{http://www.sidc.be/silso/datafiles},][]{clette16},
heliospheric modulation potential series provided by 
the Oulu cosmic ray station (\url{http://cosmicrays.oulu.fi/phi/phi.html}).
S.P. acknowledges the International Joint Research Program of ISEE, Nagoya University, 
 and thanks Naoyuki Kurita from Nagoya University for valuable discussion.
This work was partly supported by the Academy of Finland (Projects ESPERA no. 321882 and
 ReSoLVE Centre of Excellence, no. 307411).

\bibliography{bib}

\begin{thebibliography}{}

\bibitem [\protect \citeauthoryear {%
Agostinelli%
\ \protect \BOthers {.}}{%
Agostinelli%
\ \protect \BOthers {.}}{%
{\protect \APACyear {2003}}%
}]{%
Geant4-1}
\APACinsertmetastar {%
Geant4-1}%
\begin{APACrefauthors}%
Agostinelli, S.%
, Allison, J.%
, Amako, K.%
, Apostolakis, J.%
, Araujo, H.%
, Arce, P.%
\BDBL {}Zschiesche, D.%
\end{APACrefauthors}%
\unskip\
\newblock
\APACrefYearMonthDay{2003}{}{}.
\newblock
{\BBOQ}\APACrefatitle {Geant4 - a simulation toolkit} {Geant4 - a simulation
  toolkit}.{\BBCQ}
\newblock
\APACjournalVolNumPages{Nucl. Instr. Meth. Phys. A}{506}{3}{250--303}.
\newblock
\begin{APACrefURL}
  \url{http://www.sciencedirect.com/science/article/pii/S0168900203013688}
  \end{APACrefURL}
\newblock
\begin{APACrefDOI} \doi{http://dx.doi.org/10.1016/S0168-9002(03)01368-8}
  \end{APACrefDOI}
\PrintBackRefs{\CurrentBib}

\bibitem [\protect \citeauthoryear {%
Allison%
\ \protect \BOthers {.}}{%
Allison%
\ \protect \BOthers {.}}{%
{\protect \APACyear {2006}}%
}]{%
Geant4-2}
\APACinsertmetastar {%
Geant4-2}%
\begin{APACrefauthors}%
Allison, J.%
, Amako, K.%
, Apostolakis, J.%
, Araujo, H.%
, Dubois, P.%
, Asai, M.%
\BDBL {}Yoshida, H.%
\end{APACrefauthors}%
\unskip\
\newblock
\APACrefYearMonthDay{2006}{}{}.
\newblock
{\BBOQ}\APACrefatitle {Geant4 developments and applications} {Geant4
  developments and applications}.{\BBCQ}
\newblock
\APACjournalVolNumPages{Nuclear Science, IEEE Transactions on}{53}{1}{270-278}.
\newblock
\begin{APACrefDOI} \doi{10.1109/TNS.2006.869826} \end{APACrefDOI}
\PrintBackRefs{\CurrentBib}

\bibitem [\protect \citeauthoryear {%
{Asvestari}%
, {Gil}%
, {Kovaltsov}%
\BCBL {}\ \BBA {} {Usoskin}%
}{%
{Asvestari}%
\ \protect \BOthers {.}}{%
{\protect \APACyear {2017}}%
}]{%
Asvestari_JGR2017}
\APACinsertmetastar {%
Asvestari_JGR2017}%
\begin{APACrefauthors}%
{Asvestari}, E.%
, {Gil}, A.%
, {Kovaltsov}, G.%
\BCBL {}\ \BBA {} {Usoskin}, I.%
\end{APACrefauthors}%
\unskip\
\newblock
\APACrefYearMonthDay{2017}{}{}.
\newblock
{\BBOQ}\APACrefatitle {{Neutron Monitors and Cosmogenic Isotopes as Cosmic Ray
  Energy-Integration Detectors: Effective Yield Functions, Effective Energy,
  and Its Dependence on the Local Interstellar Spectrum}} {{Neutron Monitors
  and Cosmogenic Isotopes as Cosmic Ray Energy-Integration Detectors: Effective
  Yield Functions, Effective Energy, and Its Dependence on the Local
  Interstellar Spectrum}}.{\BBCQ}
\newblock
\APACjournalVolNumPages{J. Geophys. Res. Space Phys.}{122}{10}{9790-9802}.
\newblock
\begin{APACrefDOI} \doi{10.1002/2017JA024469} \end{APACrefDOI}
\PrintBackRefs{\CurrentBib}

\bibitem [\protect \citeauthoryear {%
Caballero-Lopez%
\ \BBA {} Moraal%
}{%
Caballero-Lopez%
\ \BBA {} Moraal%
}{%
{\protect \APACyear {2004}}%
}]{%
Caballero-Lopez_JGR2004}
\APACinsertmetastar {%
Caballero-Lopez_JGR2004}%
\begin{APACrefauthors}%
Caballero-Lopez, R\BPBI A.%
\BCBT {}\ \BBA {} Moraal, H.%
\end{APACrefauthors}%
\unskip\
\newblock
\APACrefYearMonthDay{2004}{}{}.
\newblock
{\BBOQ}\APACrefatitle {Limitations of the force field equation to describe
  cosmic ray modulation} {Limitations of the force field equation to describe
  cosmic ray modulation}.{\BBCQ}
\newblock
\APACjournalVolNumPages{J. Geophys. Res.: Space Phys.}{109}{A1}{A01101}.
\newblock
\begin{APACrefURL} \url{http://dx.doi.org/10.1029/2003JA010098}
  \end{APACrefURL}
\newblock
\begin{APACrefDOI} \doi{10.1029/2003JA010098} \end{APACrefDOI}
\PrintBackRefs{\CurrentBib}

\bibitem [\protect \citeauthoryear {%
{Cauquoin}%
, {Jean-Baptiste}%
, {Risi}%
, {Fourr{\'e}}%
\BCBL {}\ \BBA {} {Landais}%
}{%
{Cauquoin}%
\ \protect \BOthers {.}}{%
{\protect \APACyear {2016}}%
}]{%
cauguoin16}
\APACinsertmetastar {%
cauguoin16}%
\begin{APACrefauthors}%
{Cauquoin}, A.%
, {Jean-Baptiste}, P.%
, {Risi}, C.%
, {Fourr{\'e}}, {\'E}.%
\BCBL {}\ \BBA {} {Landais}, A.%
\end{APACrefauthors}%
\unskip\
\newblock
\APACrefYearMonthDay{2016}{}{}.
\newblock
{\BBOQ}\APACrefatitle {{Modeling the global bomb tritium transient signal with
  the AGCM LMDZ-iso: A method to evaluate aspects of the hydrological cycle}}
  {{Modeling the global bomb tritium transient signal with the AGCM LMDZ-iso: A
  method to evaluate aspects of the hydrological cycle}}.{\BBCQ}
\newblock
\APACjournalVolNumPages{J. Geophys. Res. (Atmos.)}{121}{21}{12,612-12,629}.
\newblock
\begin{APACrefDOI} \doi{10.1002/2016JD025484} \end{APACrefDOI}
\PrintBackRefs{\CurrentBib}

\bibitem [\protect \citeauthoryear {%
{Cauquoin}%
\ \protect \BOthers {.}}{%
{Cauquoin}%
\ \protect \BOthers {.}}{%
{\protect \APACyear {2015}}%
}]{%
cauquoin15}
\APACinsertmetastar {%
cauquoin15}%
\begin{APACrefauthors}%
{Cauquoin}, A.%
, {Jean-Baptiste}, P.%
, {Risi}, C.%
, {Fourr{\'e}}, {\'E}.%
, {Stenni}, B.%
\BCBL {}\ \BBA {} {Landais}, A.%
\end{APACrefauthors}%
\unskip\
\newblock
\APACrefYearMonthDay{2015}{}{}.
\newblock
{\BBOQ}\APACrefatitle {{The global distribution of natural tritium in
  precipitation simulated with an Atmospheric General Circulation Model and
  comparison with observations}} {{The global distribution of natural tritium
  in precipitation simulated with an Atmospheric General Circulation Model and
  comparison with observations}}.{\BBCQ}
\newblock
\APACjournalVolNumPages{Earth Planet. Sci. Lett.}{427}{}{160-170}.
\newblock
\begin{APACrefDOI} \doi{10.1016/j.epsl.2015.06.043} \end{APACrefDOI}
\PrintBackRefs{\CurrentBib}

\bibitem [\protect \citeauthoryear {%
{Clette}%
\ \BBA {} {Lef{\`e}vre}%
}{%
{Clette}%
\ \BBA {} {Lef{\`e}vre}%
}{%
{\protect \APACyear {2016}}%
}]{%
clette16}
\APACinsertmetastar {%
clette16}%
\begin{APACrefauthors}%
{Clette}, F.%
\BCBT {}\ \BBA {} {Lef{\`e}vre}, L.%
\end{APACrefauthors}%
\unskip\
\newblock
\APACrefYearMonthDay{2016}{}{}.
\newblock
{\BBOQ}\APACrefatitle {{The New Sunspot Number: Assembling All Corrections}}
  {{The New Sunspot Number: Assembling All Corrections}}.{\BBCQ}
\newblock
\APACjournalVolNumPages{Solar Phys.}{291}{}{2629-2651}.
\newblock
\begin{APACrefDOI} \doi{10.1007/s11207-016-1014-y} \end{APACrefDOI}
\PrintBackRefs{\CurrentBib}

\bibitem [\protect \citeauthoryear {%
{Coste}%
, {Derome}%
, {Maurin}%
\BCBL {}\ \BBA {} {Putze}%
}{%
{Coste}%
\ \protect \BOthers {.}}{%
{\protect \APACyear {2012}}%
}]{%
Coste_AnA2012}
\APACinsertmetastar {%
Coste_AnA2012}%
\begin{APACrefauthors}%
{Coste}, B.%
, {Derome}, L.%
, {Maurin}, D.%
\BCBL {}\ \BBA {} {Putze}, A.%
\end{APACrefauthors}%
\unskip\
\newblock
\APACrefYearMonthDay{2012}{}{}.
\newblock
{\BBOQ}\APACrefatitle {Constraining Galactic cosmic-ray parameters with
  $Z\leq2$~nuclei} {Constraining galactic cosmic-ray parameters with
  $z\leq2$~nuclei}.{\BBCQ}
\newblock
\APACjournalVolNumPages{Astron. Astrophys.}{539}{}{A88}.
\newblock
\begin{APACrefURL} \url{https://doi.org/10.1051/0004-6361/201117927}
  \end{APACrefURL}
\newblock
\begin{APACrefDOI} \doi{10.1051/0004-6361/201117927} \end{APACrefDOI}
\PrintBackRefs{\CurrentBib}

\bibitem [\protect \citeauthoryear {%
{Craig}%
\ \BBA {} {Lal}%
}{%
{Craig}%
\ \BBA {} {Lal}%
}{%
{\protect \APACyear {1961}}%
}]{%
craig61}
\APACinsertmetastar {%
craig61}%
\begin{APACrefauthors}%
{Craig}, H.%
\BCBT {}\ \BBA {} {Lal}, D.%
\end{APACrefauthors}%
\unskip\
\newblock
\APACrefYearMonthDay{1961}{}{}.
\newblock
{\BBOQ}\APACrefatitle {{The Production Rate of Natural Tritium}} {{The
  Production Rate of Natural Tritium}}.{\BBCQ}
\newblock
\APACjournalVolNumPages{Tellus Ser. A}{13}{1}{85-105}.
\newblock
\begin{APACrefDOI} \doi{10.1111/j.2153-3490.1961.tb00068.x} \end{APACrefDOI}
\PrintBackRefs{\CurrentBib}

\bibitem [\protect \citeauthoryear {%
{Elsasser}%
}{%
{Elsasser}%
}{%
{\protect \APACyear {1956}}%
}]{%
Elsasser_N1956}
\APACinsertmetastar {%
Elsasser_N1956}%
\begin{APACrefauthors}%
{Elsasser}, W.%
\end{APACrefauthors}%
\unskip\
\newblock
\APACrefYearMonthDay{1956}{}{}.
\newblock
{\BBOQ}\APACrefatitle {{Cosmic-Ray Intensity and Geomagnetism}} {{Cosmic-Ray
  Intensity and Geomagnetism}}.{\BBCQ}
\newblock
\APACjournalVolNumPages{Nature}{178}{}{1226-1227}.
\newblock
\begin{APACrefDOI} \doi{10.1038/1781226a0} \end{APACrefDOI}
\PrintBackRefs{\CurrentBib}

\bibitem [\protect \citeauthoryear {%
{Fireman}%
}{%
{Fireman}%
}{%
{\protect \APACyear {1953}}%
}]{%
fireman53}
\APACinsertmetastar {%
fireman53}%
\begin{APACrefauthors}%
{Fireman}, E\BPBI L.%
\end{APACrefauthors}%
\unskip\
\newblock
\APACrefYearMonthDay{1953}{}{}.
\newblock
{\BBOQ}\APACrefatitle {{Measurement of the (n, H$^{3}$) Cross Section in
  Nitrogen and Its Relationship to the Tritium Production in the Atmosphere}}
  {{Measurement of the (n, H$^{3}$) Cross Section in Nitrogen and Its
  Relationship to the Tritium Production in the Atmosphere}}.{\BBCQ}
\newblock
\APACjournalVolNumPages{Phys. Rev.}{91}{4}{922-926}.
\newblock
\begin{APACrefDOI} \doi{10.1103/PhysRev.91.922} \end{APACrefDOI}
\PrintBackRefs{\CurrentBib}

\bibitem [\protect \citeauthoryear {%
{Fourr{\'e}}%
\ \protect \BOthers {.}}{%
{Fourr{\'e}}%
\ \protect \BOthers {.}}{%
{\protect \APACyear {2018}}%
}]{%
fourre18}
\APACinsertmetastar {%
fourre18}%
\begin{APACrefauthors}%
{Fourr{\'e}}, E.%
, {Landais}, A.%
, {Cauquoin}, A.%
, {Jean-Baptiste}, P.%
, {Lipenkov}, V.%
\BCBL {}\ \BBA {} {Petit}, J\BPBI R.%
\end{APACrefauthors}%
\unskip\
\newblock
\APACrefYearMonthDay{2018}{}{}.
\newblock
{\BBOQ}\APACrefatitle {{Tritium Records to Trace Stratospheric Moisture Inputs
  in Antarctica}} {{Tritium Records to Trace Stratospheric Moisture Inputs in
  Antarctica}}.{\BBCQ}
\newblock
\APACjournalVolNumPages{J. Geophys. Res. (Atmos.)}{123}{6}{3009-3018}.
\newblock
\begin{APACrefDOI} \doi{10.1002/2018JD028304} \end{APACrefDOI}
\PrintBackRefs{\CurrentBib}

\bibitem [\protect \citeauthoryear {%
{Geant4 collaboration}%
}{%
{Geant4 collaboration}%
}{%
{\protect \APACyear {2013}}%
}]{%
Geant4-Phys2013}
\APACinsertmetastar {%
Geant4-Phys2013}%
\begin{APACrefauthors}%
{Geant4 collaboration}.%
\end{APACrefauthors}%
\unskip\
\newblock
\APACrefYearMonthDay{2013}{}{}.
\newblock
{\BBOQ}\APACrefatitle {Physics reference manual (version geant4 9.10.0)}
  {Physics reference manual (version geant4 9.10.0)}{\BBCQ}\
  [\bibcomputersoftwaremanual].
\newblock
\APACrefnote{available from \texttt{http://geant4.cern.ch/support/index.shtml}}
\PrintBackRefs{\CurrentBib}

\bibitem [\protect \citeauthoryear {%
{Gleeson}%
\ \BBA {} {Axford}%
}{%
{Gleeson}%
\ \BBA {} {Axford}%
}{%
{\protect \APACyear {1967}}%
}]{%
Gleeson_ApJ1967}
\APACinsertmetastar {%
Gleeson_ApJ1967}%
\begin{APACrefauthors}%
{Gleeson}, J\BPBI J.%
\BCBT {}\ \BBA {} {Axford}, W\BPBI I.%
\end{APACrefauthors}%
\unskip\
\newblock
\APACrefYearMonthDay{1967}{}{}.
\newblock
{\BBOQ}\APACrefatitle {{Cosmic rays in the interplanetary medium}} {{Cosmic
  rays in the interplanetary medium}}.{\BBCQ}
\newblock
\APACjournalVolNumPages{Astrophys. J.}{149}{}{L115}.
\newblock
\begin{APACrefURL} \url{http://adsabs.harvard.edu/abs/1967ApJ...149L.115G}
  \end{APACrefURL}
\newblock
\begin{APACrefDOI} \doi{10.1086/180070} \end{APACrefDOI}
\PrintBackRefs{\CurrentBib}

\bibitem [\protect \citeauthoryear {%
{Grieder}%
}{%
{Grieder}%
}{%
{\protect \APACyear {2001}}%
}]{%
Grieder2001}
\APACinsertmetastar {%
Grieder2001}%
\begin{APACrefauthors}%
{Grieder}, P\BPBI K\BPBI F.%
\end{APACrefauthors}%
\unskip\
\newblock
\APACrefYear{2001}.
\newblock
\APACrefbtitle {{Cosmic Rays at Earth}} {{Cosmic Rays at Earth}}.
\newblock
\APACaddressPublisher{Amsterdam}{Elsevier Science}.
\PrintBackRefs{\CurrentBib}

\bibitem [\protect \citeauthoryear {%
Herbst%
\ \protect \BOthers {.}}{%
Herbst%
\ \protect \BOthers {.}}{%
{\protect \APACyear {2010}}%
}]{%
Herbst_JGR2010}
\APACinsertmetastar {%
Herbst_JGR2010}%
\begin{APACrefauthors}%
Herbst, K.%
, Kopp, A.%
, Heber, B.%
, Steinhilber, F.%
, Fichtner, H.%
, Scherer, K.%
\BCBL {}\ \BBA {} Matthiä, D.%
\end{APACrefauthors}%
\unskip\
\newblock
\APACrefYearMonthDay{2010}{}{}.
\newblock
{\BBOQ}\APACrefatitle {On the importance of the local interstellar spectrum for
  the solar modulation parameter} {On the importance of the local interstellar
  spectrum for the solar modulation parameter}.{\BBCQ}
\newblock
\APACjournalVolNumPages{J. Geophys. Res.: Atmos.}{115}{D1}{D00I20}.
\newblock
\begin{APACrefURL} \url{http://dx.doi.org/10.1029/2009JD012557}
  \end{APACrefURL}
\newblock
\begin{APACrefDOI} \doi{10.1029/2009JD012557} \end{APACrefDOI}
\PrintBackRefs{\CurrentBib}

\bibitem [\protect \citeauthoryear {%
Herbst%
, Muscheler%
\BCBL {}\ \BBA {} Heber%
}{%
Herbst%
\ \protect \BOthers {.}}{%
{\protect \APACyear {2017}}%
}]{%
Herbst_JGR2017}
\APACinsertmetastar {%
Herbst_JGR2017}%
\begin{APACrefauthors}%
Herbst, K.%
, Muscheler, R.%
\BCBL {}\ \BBA {} Heber, B.%
\end{APACrefauthors}%
\unskip\
\newblock
\APACrefYearMonthDay{2017}{}{}.
\newblock
{\BBOQ}\APACrefatitle {The new local interstellar spectra and their influence
  on the production rates of the cosmogenic radionuclides {$^{10}$Be} and
  {$^{14}$C}} {The new local interstellar spectra and their influence on the
  production rates of the cosmogenic radionuclides {$^{10}$Be} and
  {$^{14}$C}}.{\BBCQ}
\newblock
\APACjournalVolNumPages{Journal of Geophysical Research: Space
  Physics}{122}{1}{23-34}.
\newblock
\begin{APACrefDOI} \doi{10.1002/2016JA023207} \end{APACrefDOI}
\PrintBackRefs{\CurrentBib}

\bibitem [\protect \citeauthoryear {%
{Juhlke}%
\ \protect \BOthers {.}}{%
{Juhlke}%
\ \protect \BOthers {.}}{%
{\protect \APACyear {2020}}%
}]{%
juhlke20}
\APACinsertmetastar {%
juhlke20}%
\begin{APACrefauthors}%
{Juhlke}, T.%
, {S{\"u}ltenfu{\ss}}, J.%
, {Trachte}, K.%
, {Huneau}, F.%
, {Garel}, E.%
, {Santoni}, S.%
\BDBL {}{van Geldern}, R.%
\end{APACrefauthors}%
\unskip\
\newblock
\APACrefYearMonthDay{2020}{}{}.
\newblock
{\BBOQ}\APACrefatitle {{Tritium as a hydrological tracer in Mediterranean
  precipitation events}} {{Tritium as a hydrological tracer in Mediterranean
  precipitation events}}.{\BBCQ}
\newblock
\APACjournalVolNumPages{Atmos. Chem. Phys.}{20}{6}{3555-3568}.
\newblock
\begin{APACrefDOI} \doi{10.5194/acp-20-3555-2020} \end{APACrefDOI}
\PrintBackRefs{\CurrentBib}

\bibitem [\protect \citeauthoryear {%
{Koldobskiy}%
, {Bindi}%
, {Corti}%
, {Kovaltsov}%
\BCBL {}\ \BBA {} {Usoskin}%
}{%
{Koldobskiy}%
\ \protect \BOthers {.}}{%
{\protect \APACyear {2019}}%
}]{%
Koldobskiy_JGR2019}
\APACinsertmetastar {%
Koldobskiy_JGR2019}%
\begin{APACrefauthors}%
{Koldobskiy}, S\BPBI A.%
, {Bindi}, V.%
, {Corti}, C.%
, {Kovaltsov}, G\BPBI A.%
\BCBL {}\ \BBA {} {Usoskin}, I\BPBI G.%
\end{APACrefauthors}%
\unskip\
\newblock
\APACrefYearMonthDay{2019}{}{}.
\newblock
{\BBOQ}\APACrefatitle {{Validation of the Neutron Monitor Yield Function Using
  Data From AMS-02 Experiment, 2011-2017}} {{Validation of the Neutron Monitor
  Yield Function Using Data From AMS-02 Experiment, 2011-2017}}.{\BBCQ}
\newblock
\APACjournalVolNumPages{J. Geophys. Res.: Space Phys.}{124}{4}{2367-2379}.
\newblock
\begin{APACrefDOI} \doi{10.1029/2018JA026340} \end{APACrefDOI}
\PrintBackRefs{\CurrentBib}

\bibitem [\protect \citeauthoryear {%
{Kovaltsov}%
, {Mishev}%
\BCBL {}\ \BBA {} {Usoskin}%
}{%
{Kovaltsov}%
\ \protect \BOthers {.}}{%
{\protect \APACyear {2012}}%
}]{%
Kovaltsov_EPSL2012}
\APACinsertmetastar {%
Kovaltsov_EPSL2012}%
\begin{APACrefauthors}%
{Kovaltsov}, G\BPBI A.%
, {Mishev}, A.%
\BCBL {}\ \BBA {} {Usoskin}, I\BPBI G.%
\end{APACrefauthors}%
\unskip\
\newblock
\APACrefYearMonthDay{2012}{}{}.
\newblock
{\BBOQ}\APACrefatitle {{A new model of cosmogenic production of radiocarbon
  $^{14}$C in the atmosphere}} {{A new model of cosmogenic production of
  radiocarbon $^{14}$C in the atmosphere}}.{\BBCQ}
\newblock
\APACjournalVolNumPages{Earth Planet. Sci. Lett.}{337}{}{114-120}.
\newblock
\begin{APACrefDOI} \doi{10.1016/j.epsl.2012.05.036} \end{APACrefDOI}
\PrintBackRefs{\CurrentBib}

\bibitem [\protect \citeauthoryear {%
{Kovaltsov}%
\ \BBA {} {Usoskin}%
}{%
{Kovaltsov}%
\ \BBA {} {Usoskin}%
}{%
{\protect \APACyear {2010}}%
}]{%
Kovaltsov_EPSL2010}
\APACinsertmetastar {%
Kovaltsov_EPSL2010}%
\begin{APACrefauthors}%
{Kovaltsov}, G\BPBI A.%
\BCBT {}\ \BBA {} {Usoskin}, I\BPBI G.%
\end{APACrefauthors}%
\unskip\
\newblock
\APACrefYearMonthDay{2010}{}{}.
\newblock
{\BBOQ}\APACrefatitle {{A new 3D numerical model of cosmogenic nuclide
  $^{10}$Be production in the atmosphere}} {{A new 3D numerical model of
  cosmogenic nuclide $^{10}$Be production in the atmosphere}}.{\BBCQ}
\newblock
\APACjournalVolNumPages{Earth Planet. Sci. Lett.}{291}{}{182-188}.
\newblock
\begin{APACrefDOI} \doi{10.1016/j.epsl.2010.01.011} \end{APACrefDOI}
\PrintBackRefs{\CurrentBib}

\bibitem [\protect \citeauthoryear {%
Lal%
\ \BBA {} Peters%
}{%
Lal%
\ \BBA {} Peters%
}{%
{\protect \APACyear {1967}}%
}]{%
lal67}
\APACinsertmetastar {%
lal67}%
\begin{APACrefauthors}%
Lal, D.%
\BCBT {}\ \BBA {} Peters, B.%
\end{APACrefauthors}%
\unskip\
\newblock
\APACrefYearMonthDay{1967}{}{}.
\newblock
{\BBOQ}\APACrefatitle {Cosmic Ray Produced Radioactivity on the Earth} {Cosmic
  ray produced radioactivity on the earth}.{\BBCQ}
\newblock
\BIn{} K.~Sittle\ (\BED), \APACrefbtitle {Handbuch der Physik} {Handbuch der
  physik}\ (\BVOL~46, \BPGS\ 551--612).
\newblock
\APACaddressPublisher{Berlin}{Springer}.
\PrintBackRefs{\CurrentBib}

\bibitem [\protect \citeauthoryear {%
{L{\'a}szl{\'o}}%
, {Palcsu}%
\BCBL {}\ \BBA {} {Leel\H{o}ssy}%
}{%
{L{\'a}szl{\'o}}%
\ \protect \BOthers {.}}{%
{\protect \APACyear {2020}}%
}]{%
laszlo20}
\APACinsertmetastar {%
laszlo20}%
\begin{APACrefauthors}%
{L{\'a}szl{\'o}}, E.%
, {Palcsu}, M.%
\BCBL {}\ \BBA {} {Leel\H{o}ssy}, {\'A}.%
\end{APACrefauthors}%
\unskip\
\newblock
\APACrefYearMonthDay{2020}{}{}.
\newblock
{\BBOQ}\APACrefatitle {{Estimation of the solar-induced natural variability of
  the tritium concentration of precipitation in the Northern and Southern
  Hemisphere}} {{Estimation of the solar-induced natural variability of the
  tritium concentration of precipitation in the Northern and Southern
  Hemisphere}}.{\BBCQ}
\newblock
\APACjournalVolNumPages{Atmos. Envir.}{}{}{}.
\newblock
\begin{APACrefDOI} \doi{10.1016/j.atmosenv.2020.117605} \end{APACrefDOI}
\PrintBackRefs{\CurrentBib}

\bibitem [\protect \citeauthoryear {%
{Masarik}%
\ \BBA {} {Beer}%
}{%
{Masarik}%
\ \BBA {} {Beer}%
}{%
{\protect \APACyear {1999}}%
}]{%
Masarik_JGR1999}
\APACinsertmetastar {%
Masarik_JGR1999}%
\begin{APACrefauthors}%
{Masarik}, J.%
\BCBT {}\ \BBA {} {Beer}, J.%
\end{APACrefauthors}%
\unskip\
\newblock
\APACrefYearMonthDay{1999}{}{}.
\newblock
{\BBOQ}\APACrefatitle {{Simulation of particle fluxes and cosmogenic nuclide
  production in the Earth's atmosphere}} {{Simulation of particle fluxes and
  cosmogenic nuclide production in the Earth's atmosphere}}.{\BBCQ}
\newblock
\APACjournalVolNumPages{J. Geophys. Res.}{104}{}{12099-12112}.
\newblock
\begin{APACrefDOI} \doi{10.1029/1998JD200091} \end{APACrefDOI}
\PrintBackRefs{\CurrentBib}

\bibitem [\protect \citeauthoryear {%
{Masarik}%
\ \BBA {} {Beer}%
}{%
{Masarik}%
\ \BBA {} {Beer}%
}{%
{\protect \APACyear {2009}}%
}]{%
Masarik_JGR2009}
\APACinsertmetastar {%
Masarik_JGR2009}%
\begin{APACrefauthors}%
{Masarik}, J.%
\BCBT {}\ \BBA {} {Beer}, J.%
\end{APACrefauthors}%
\unskip\
\newblock
\APACrefYearMonthDay{2009}{}{}.
\newblock
{\BBOQ}\APACrefatitle {{An updated simulation of particle fluxes and cosmogenic
  nuclide production in the Earth's atmosphere}} {{An updated simulation of
  particle fluxes and cosmogenic nuclide production in the Earth's
  atmosphere}}.{\BBCQ}
\newblock
\APACjournalVolNumPages{J. Geophys. Res.}{114}{}{{D11103}}.
\newblock
\begin{APACrefDOI} \doi{10.1029/2008JD010557} \end{APACrefDOI}
\PrintBackRefs{\CurrentBib}

\bibitem [\protect \citeauthoryear {%
{Masarik}%
\ \BBA {} {Reedy}%
}{%
{Masarik}%
\ \BBA {} {Reedy}%
}{%
{\protect \APACyear {1995}}%
}]{%
masarik95}
\APACinsertmetastar {%
masarik95}%
\begin{APACrefauthors}%
{Masarik}, J.%
\BCBT {}\ \BBA {} {Reedy}, R\BPBI C.%
\end{APACrefauthors}%
\unskip\
\newblock
\APACrefYearMonthDay{1995}{}{}.
\newblock
{\BBOQ}\APACrefatitle {{Terrestrial cosmogenic-nuclide production systematics
  calculated from numerical simulations}} {{Terrestrial cosmogenic-nuclide
  production systematics calculated from numerical simulations}}.{\BBCQ}
\newblock
\APACjournalVolNumPages{Earth Planet. Sci. Lett.}{136}{}{381-395}.
\newblock
\begin{APACrefDOI} \doi{10.1016/0012-821X(95)00169-D} \end{APACrefDOI}
\PrintBackRefs{\CurrentBib}

\bibitem [\protect \citeauthoryear {%
{Mesick}%
, {Feldman}%
, {Coupland}%
\BCBL {}\ \BBA {} {Stonehill}%
}{%
{Mesick}%
\ \protect \BOthers {.}}{%
{\protect \APACyear {2018}}%
}]{%
mesick18}
\APACinsertmetastar {%
mesick18}%
\begin{APACrefauthors}%
{Mesick}, K\BPBI E.%
, {Feldman}, W\BPBI C.%
, {Coupland}, D\BPBI D\BPBI S.%
\BCBL {}\ \BBA {} {Stonehill}, L\BPBI C.%
\end{APACrefauthors}%
\unskip\
\newblock
\APACrefYearMonthDay{2018}{}{}.
\newblock
{\BBOQ}\APACrefatitle {{Benchmarking Geant4 for Simulating Galactic Cosmic Ray
  Interactions Within Planetary Bodies}} {{Benchmarking Geant4 for Simulating
  Galactic Cosmic Ray Interactions Within Planetary Bodies}}.{\BBCQ}
\newblock
\APACjournalVolNumPages{Earth Space Sci.}{5}{7}{324-338}.
\newblock
\begin{APACrefDOI} \doi{10.1029/2018EA000400} \end{APACrefDOI}
\PrintBackRefs{\CurrentBib}

\bibitem [\protect \citeauthoryear {%
{Michel}%
}{%
{Michel}%
}{%
{\protect \APACyear {2005}}%
}]{%
michel05}
\APACinsertmetastar {%
michel05}%
\begin{APACrefauthors}%
{Michel}, R.%
\end{APACrefauthors}%
\unskip\
\newblock
\APACrefYearMonthDay{2005}{}{}.
\newblock
{\BBOQ}\APACrefatitle {{Tritium in the hydrological cycle}} {{Tritium in the
  hydrological cycle}}.{\BBCQ}
\newblock
\BIn{} P.~{Aggarwal}, J.~{Gat}\BCBL {}\ \BBA {} K.~{Froehlich}\ (\BEDS),
  \APACrefbtitle {{Isotopes in the Water Cycle. Past, Present and Future of a
  Developing Science}} {{Isotopes in the Water Cycle. Past, Present and Future
  of a Developing Science}}\ (\BPG~53-66).
\newblock
\APACaddressPublisher{Dordtrecht}{Springer}.
\PrintBackRefs{\CurrentBib}

\bibitem [\protect \citeauthoryear {%
Mishev%
, Kocharov%
\BCBL {}\ \BBA {} Usoskin%
}{%
Mishev%
\ \protect \BOthers {.}}{%
{\protect \APACyear {2014}}%
}]{%
Mishev2014}
\APACinsertmetastar {%
Mishev2014}%
\begin{APACrefauthors}%
Mishev, A\BPBI L.%
, Kocharov, L\BPBI G.%
\BCBL {}\ \BBA {} Usoskin, I\BPBI G.%
\end{APACrefauthors}%
\unskip\
\newblock
\APACrefYearMonthDay{2014}{}{}.
\newblock
{\BBOQ}\APACrefatitle {Analysis of the ground level enhancement on 17 May 2012
  using data from the global neutron monitor network} {Analysis of the ground
  level enhancement on 17 may 2012 using data from the global neutron monitor
  network}.{\BBCQ}
\newblock
\APACjournalVolNumPages{J. Geophys. Res.: Space Phys.}{119}{2}{670--679}.
\newblock
\begin{APACrefURL} \url{http://dx.doi.org/10.1002/2013JA019253}
  \end{APACrefURL}
\newblock
\begin{APACrefDOI} \doi{10.1002/2013JA019253} \end{APACrefDOI}
\PrintBackRefs{\CurrentBib}

\bibitem [\protect \citeauthoryear {%
Nevalainen%
, Usoskin%
\BCBL {}\ \BBA {} Mishev%
}{%
Nevalainen%
\ \protect \BOthers {.}}{%
{\protect \APACyear {2013}}%
}]{%
Nevalainen2013}
\APACinsertmetastar {%
Nevalainen2013}%
\begin{APACrefauthors}%
Nevalainen, J.%
, Usoskin, I.%
\BCBL {}\ \BBA {} Mishev, A.%
\end{APACrefauthors}%
\unskip\
\newblock
\APACrefYearMonthDay{2013}{}{}.
\newblock
{\BBOQ}\APACrefatitle {Eccentric dipole approximation of the geomagnetic field:
  Application to cosmic ray computations} {Eccentric dipole approximation of
  the geomagnetic field: Application to cosmic ray computations}.{\BBCQ}
\newblock
\APACjournalVolNumPages{Adv. Space Res.}{52}{1}{22-29}.
\newblock
\begin{APACrefDOI} \doi{http://dx.doi.org/10.1016/j.asr.2013.02.020}
  \end{APACrefDOI}
\PrintBackRefs{\CurrentBib}

\bibitem [\protect \citeauthoryear {%
{Nir}%
, {Kruger}%
, {Lingenfelter}%
\BCBL {}\ \BBA {} {Flamm}%
}{%
{Nir}%
\ \protect \BOthers {.}}{%
{\protect \APACyear {1966}}%
}]{%
Nir_RGSP1966}
\APACinsertmetastar {%
Nir_RGSP1966}%
\begin{APACrefauthors}%
{Nir}, A.%
, {Kruger}, S\BPBI T.%
, {Lingenfelter}, R\BPBI E.%
\BCBL {}\ \BBA {} {Flamm}, E\BPBI J.%
\end{APACrefauthors}%
\unskip\
\newblock
\APACrefYearMonthDay{1966}{}{}.
\newblock
{\BBOQ}\APACrefatitle {{Natural Tritium}} {{Natural Tritium}}.{\BBCQ}
\newblock
\APACjournalVolNumPages{Rev. Geophys. Space Phys.}{4}{}{441-456}.
\newblock
\begin{APACrefDOI} \doi{10.1029/RG004i004p00441} \end{APACrefDOI}
\PrintBackRefs{\CurrentBib}

\bibitem [\protect \citeauthoryear {%
{O'Brien}%
}{%
{O'Brien}%
}{%
{\protect \APACyear {1979}}%
}]{%
OBrien_JGR1979}
\APACinsertmetastar {%
OBrien_JGR1979}%
\begin{APACrefauthors}%
{O'Brien}, K.%
\end{APACrefauthors}%
\unskip\
\newblock
\APACrefYearMonthDay{1979}{}{}.
\newblock
{\BBOQ}\APACrefatitle {{Secular variations in the production of cosmogenic
  isotopes in the earth's atmosphere}} {{Secular variations in the production
  of cosmogenic isotopes in the earth's atmosphere}}.{\BBCQ}
\newblock
\APACjournalVolNumPages{J. Geophys. Res.}{84}{}{423-431}.
\newblock
\begin{APACrefDOI} \doi{10.1029/JA084iA02p00423} \end{APACrefDOI}
\PrintBackRefs{\CurrentBib}

\bibitem [\protect \citeauthoryear {%
{Palcsu}%
\ \protect \BOthers {.}}{%
{Palcsu}%
\ \protect \BOthers {.}}{%
{\protect \APACyear {2018}}%
}]{%
palcsu18}
\APACinsertmetastar {%
palcsu18}%
\begin{APACrefauthors}%
{Palcsu}, L.%
, {Morgenstern}, U.%
, {S{\"u}ltenfuss}, J.%
, {Koltai}, G.%
, {L{\'a}szl{\'o}}, E.%
, {Temovski}, M.%
\BDBL {}{Jull}, A\BPBI J.%
\end{APACrefauthors}%
\unskip\
\newblock
\APACrefYearMonthDay{2018}{}{}.
\newblock
{\BBOQ}\APACrefatitle {{Modulation of Cosmogenic Tritium in Meteoric
  Precipitation by the 11-year Cycle of Solar Magnetic Field Activity}}
  {{Modulation of Cosmogenic Tritium in Meteoric Precipitation by the 11-year
  Cycle of Solar Magnetic Field Activity}}.{\BBCQ}
\newblock
\APACjournalVolNumPages{Sci. Rep.}{8}{}{12813}.
\newblock
\begin{APACrefDOI} \doi{10.1038/s41598-018-31208-9} \end{APACrefDOI}
\PrintBackRefs{\CurrentBib}

\bibitem [\protect \citeauthoryear {%
Picone%
, Hedin%
, Drob%
\BCBL {}\ \BBA {} Aikin%
}{%
Picone%
\ \protect \BOthers {.}}{%
{\protect \APACyear {2002}}%
}]{%
Picone2002}
\APACinsertmetastar {%
Picone2002}%
\begin{APACrefauthors}%
Picone, J\BPBI M.%
, Hedin, A\BPBI E.%
, Drob, D\BPBI P.%
\BCBL {}\ \BBA {} Aikin, A\BPBI C.%
\end{APACrefauthors}%
\unskip\
\newblock
\APACrefYearMonthDay{2002}{}{}.
\newblock
{\BBOQ}\APACrefatitle {NRLMSISE-00 empirical model of the atmosphere:
  Statistical comparisons and scientific issues} {Nrlmsise-00 empirical model
  of the atmosphere: Statistical comparisons and scientific issues}.{\BBCQ}
\newblock
\APACjournalVolNumPages{J. Geophys. Res.: Space Phys.}{107}{A12}{SIA 15-1--SIA
  15-16}.
\newblock
\begin{APACrefURL} \url{http://dx.doi.org/10.1029/2002JA009430}
  \end{APACrefURL}
\newblock
\APACrefnote{1468}
\newblock
\begin{APACrefDOI} \doi{10.1029/2002JA009430} \end{APACrefDOI}
\PrintBackRefs{\CurrentBib}

\bibitem [\protect \citeauthoryear {%
{Poluianov}%
, {Kovaltsov}%
, {Mishev}%
\BCBL {}\ \BBA {} {Usoskin}%
}{%
{Poluianov}%
\ \protect \BOthers {.}}{%
{\protect \APACyear {2016}}%
}]{%
Poluianov_JGR2016}
\APACinsertmetastar {%
Poluianov_JGR2016}%
\begin{APACrefauthors}%
{Poluianov}, S\BPBI V.%
, {Kovaltsov}, G\BPBI A.%
, {Mishev}, A\BPBI L.%
\BCBL {}\ \BBA {} {Usoskin}, I\BPBI G.%
\end{APACrefauthors}%
\unskip\
\newblock
\APACrefYearMonthDay{2016}{}{}.
\newblock
{\BBOQ}\APACrefatitle {{Production of cosmogenic isotopes $^{7}$Be, $^{10}$Be,
  $^{14}$C, $^{22}$Na, and $^{36}$Cl in the atmosphere: Altitudinal profiles of
  yield functions}} {{Production of cosmogenic isotopes $^{7}$Be, $^{10}$Be,
  $^{14}$C, $^{22}$Na, and $^{36}$Cl in the atmosphere: Altitudinal profiles of
  yield functions}}.{\BBCQ}
\newblock
\APACjournalVolNumPages{Journal of Geophysical Research
  (Atmospheres)}{121}{}{8125-8136}.
\newblock
\begin{APACrefDOI} \doi{10.1002/2016JD025034} \end{APACrefDOI}
\PrintBackRefs{\CurrentBib}

\bibitem [\protect \citeauthoryear {%
{Raukunen}%
\ \protect \BOthers {.}}{%
{Raukunen}%
\ \protect \BOthers {.}}{%
{\protect \APACyear {2018}}%
}]{%
Raukunen_SWSC2018}
\APACinsertmetastar {%
Raukunen_SWSC2018}%
\begin{APACrefauthors}%
{Raukunen}, O.%
, {Vainio}, R.%
, {Tylka}, A\BPBI J.%
, {Dietrich}, W\BPBI F.%
, {Jiggens}, P.%
, {Heynderickx}, D.%
\BDBL {}{Siipola}, R.%
\end{APACrefauthors}%
\unskip\
\newblock
\APACrefYearMonthDay{2018}{}{}.
\newblock
{\BBOQ}\APACrefatitle {{Two solar proton fluence models based on ground level
  enhancement observations}} {{Two solar proton fluence models based on ground
  level enhancement observations}}.{\BBCQ}
\newblock
\APACjournalVolNumPages{Journal of Space Weather and Space
  Climate}{8}{27}{A04}.
\newblock
\begin{APACrefDOI} \doi{10.1051/swsc/2017031} \end{APACrefDOI}
\PrintBackRefs{\CurrentBib}

\bibitem [\protect \citeauthoryear {%
Smart%
\ \BBA {} Shea%
}{%
Smart%
\ \BBA {} Shea%
}{%
{\protect \APACyear {2009}}%
}]{%
Smart2009}
\APACinsertmetastar {%
Smart2009}%
\begin{APACrefauthors}%
Smart, D.%
\BCBT {}\ \BBA {} Shea, M.%
\end{APACrefauthors}%
\unskip\
\newblock
\APACrefYearMonthDay{2009}{}{}.
\newblock
{\BBOQ}\APACrefatitle {Fifty years of progress in geomagnetic cutoff rigidity
  determinations} {Fifty years of progress in geomagnetic cutoff rigidity
  determinations}.{\BBCQ}
\newblock
\APACjournalVolNumPages{Adv. Space Res.}{44}{10}{1107--1123}.
\newblock
\begin{APACrefURL}
  \url{http://www.sciencedirect.com/science/article/pii/S0273117709004815}
  \end{APACrefURL}
\newblock
\begin{APACrefDOI} \doi{http://dx.doi.org/10.1016/j.asr.2009.07.005}
  \end{APACrefDOI}
\PrintBackRefs{\CurrentBib}

\bibitem [\protect \citeauthoryear {%
{Smart}%
, {Shea}%
\BCBL {}\ \BBA {} {Fl{\"u}ckiger}%
}{%
{Smart}%
\ \protect \BOthers {.}}{%
{\protect \APACyear {2000}}%
}]{%
shea00}
\APACinsertmetastar {%
shea00}%
\begin{APACrefauthors}%
{Smart}, D\BPBI F.%
, {Shea}, M\BPBI A.%
\BCBL {}\ \BBA {} {Fl{\"u}ckiger}, E\BPBI O.%
\end{APACrefauthors}%
\unskip\
\newblock
\APACrefYearMonthDay{2000}{}{}.
\newblock
{\BBOQ}\APACrefatitle {{Magnetospheric Models and Trajectory Computations}}
  {{Magnetospheric Models and Trajectory Computations}}.{\BBCQ}
\newblock
\APACjournalVolNumPages{Space Sci. Rev.}{93}{}{305-333}.
\newblock
\begin{APACrefDOI} \doi{10.1023/A:1026556831199} \end{APACrefDOI}
\PrintBackRefs{\CurrentBib}

\bibitem [\protect \citeauthoryear {%
{Sykora}%
\ \BBA {} {Froehlich}%
}{%
{Sykora}%
\ \BBA {} {Froehlich}%
}{%
{\protect \APACyear {2010}}%
}]{%
froehlich10}
\APACinsertmetastar {%
froehlich10}%
\begin{APACrefauthors}%
{Sykora}, I.%
\BCBT {}\ \BBA {} {Froehlich}, K.%
\end{APACrefauthors}%
\unskip\
\newblock
\APACrefYearMonthDay{2010}{}{}.
\newblock
{\BBOQ}\APACrefatitle {{Radionuclides as Tracers of Atmospheric Processes}}
  {{Radionuclides as Tracers of Atmospheric Processes}}.{\BBCQ}
\newblock
\BIn{} K.~{Froehlich}\ (\BED), \APACrefbtitle {{Environmental Radionuclides:
  Tracers and Timers of Terrestrial Processes}} {{Environmental Radionuclides:
  Tracers and Timers of Terrestrial Processes}}\ (\BVOL~16, \BPG~51-88).
\newblock
\APACaddressPublisher{Amsterdam}{Elevier}.
\PrintBackRefs{\CurrentBib}

\bibitem [\protect \citeauthoryear {%
{Tatischeff}%
, {Kozlovsky}%
, {Kiener}%
\BCBL {}\ \BBA {} {Murphy}%
}{%
{Tatischeff}%
\ \protect \BOthers {.}}{%
{\protect \APACyear {2006}}%
}]{%
Tatischeff_APJS2006}
\APACinsertmetastar {%
Tatischeff_APJS2006}%
\begin{APACrefauthors}%
{Tatischeff}, V.%
, {Kozlovsky}, B.%
, {Kiener}, J.%
\BCBL {}\ \BBA {} {Murphy}, R\BPBI J.%
\end{APACrefauthors}%
\unskip\
\newblock
\APACrefYearMonthDay{2006}{}{}.
\newblock
{\BBOQ}\APACrefatitle {{Delayed X- and Gamma-Ray Line Emission from Solar Flare
  Radioactivity}} {{Delayed X- and Gamma-Ray Line Emission from Solar Flare
  Radioactivity}}.{\BBCQ}
\newblock
\APACjournalVolNumPages{Astrophys. J. Suppl.}{165}{}{606-617}.
\newblock
\begin{APACrefDOI} \doi{10.1086/505112} \end{APACrefDOI}
\PrintBackRefs{\CurrentBib}

\bibitem [\protect \citeauthoryear {%
{Th{\'e}bault}%
\ \protect \BOthers {.}}{%
{Th{\'e}bault}%
\ \protect \BOthers {.}}{%
{\protect \APACyear {2015}}%
}]{%
Thebault_EPS2015}
\APACinsertmetastar {%
Thebault_EPS2015}%
\begin{APACrefauthors}%
{Th{\'e}bault}, E.%
, {Finlay}, C\BPBI C.%
, {Beggan}, C\BPBI D.%
, {Alken}, P.%
, {Aubert}, J.%
, {Barrois}, O.%
\BDBL {}{Zvereva}, T.%
\end{APACrefauthors}%
\unskip\
\newblock
\APACrefYearMonthDay{2015}{}{}.
\newblock
{\BBOQ}\APACrefatitle {{International Geomagnetic Reference Field: the 12th
  generation}} {{International Geomagnetic Reference Field: the 12th
  generation}}.{\BBCQ}
\newblock
\APACjournalVolNumPages{Earth Planet. Space}{67}{}{79}.
\newblock
\begin{APACrefDOI} \doi{10.1186/s40623-015-0228-9} \end{APACrefDOI}
\PrintBackRefs{\CurrentBib}

\bibitem [\protect \citeauthoryear {%
{Usoskin}%
, {Alanko-Huotari}%
, {Kovaltsov}%
\BCBL {}\ \BBA {} {Mursula}%
}{%
{Usoskin}%
\ \protect \BOthers {.}}{%
{\protect \APACyear {2005}}%
}]{%
Usoskin_phi_05}
\APACinsertmetastar {%
Usoskin_phi_05}%
\begin{APACrefauthors}%
{Usoskin}, I\BPBI G.%
, {Alanko-Huotari}, K.%
, {Kovaltsov}, G\BPBI A.%
\BCBL {}\ \BBA {} {Mursula}, K.%
\end{APACrefauthors}%
\unskip\
\newblock
\APACrefYearMonthDay{2005}{}{}.
\newblock
{\BBOQ}\APACrefatitle {{Heliospheric modulation of cosmic rays: Monthly
  reconstruction for 1951--2004}} {{Heliospheric modulation of cosmic rays:
  Monthly reconstruction for 1951--2004}}.{\BBCQ}
\newblock
\APACjournalVolNumPages{J. Geophys. Res.}{110}{}{{A12108}}.
\newblock
\begin{APACrefDOI} \doi{10.1029/2005JA011250} \end{APACrefDOI}
\PrintBackRefs{\CurrentBib}

\bibitem [\protect \citeauthoryear {%
{Usoskin}%
, {Gil}%
, {Kovaltsov}%
, {Mishev}%
\BCBL {}\ \BBA {} {Mikhailov}%
}{%
{Usoskin}%
\ \protect \BOthers {.}}{%
{\protect \APACyear {2017}}%
}]{%
Usoskin_JGR2017}
\APACinsertmetastar {%
Usoskin_JGR2017}%
\begin{APACrefauthors}%
{Usoskin}, I\BPBI G.%
, {Gil}, A.%
, {Kovaltsov}, G\BPBI A.%
, {Mishev}, A\BPBI L.%
\BCBL {}\ \BBA {} {Mikhailov}, V\BPBI V.%
\end{APACrefauthors}%
\unskip\
\newblock
\APACrefYearMonthDay{2017}{}{}.
\newblock
{\BBOQ}\APACrefatitle {{Heliospheric modulation of cosmic rays during the
  neutron monitor era: Calibration using PAMELA data for 2006-2010}}
  {{Heliospheric modulation of cosmic rays during the neutron monitor era:
  Calibration using PAMELA data for 2006-2010}}.{\BBCQ}
\newblock
\APACjournalVolNumPages{J. Geophys. Res.}{122}{}{3875-3887}.
\newblock
\begin{APACrefDOI} \doi{10.1002/2016JA023819} \end{APACrefDOI}
\PrintBackRefs{\CurrentBib}

\bibitem [\protect \citeauthoryear {%
{Usoskin}%
\ \BBA {} {Kovaltsov}%
}{%
{Usoskin}%
\ \BBA {} {Kovaltsov}%
}{%
{\protect \APACyear {2008}}%
}]{%
Usoskin_JGR2008}
\APACinsertmetastar {%
Usoskin_JGR2008}%
\begin{APACrefauthors}%
{Usoskin}, I\BPBI G.%
\BCBT {}\ \BBA {} {Kovaltsov}, G\BPBI A.%
\end{APACrefauthors}%
\unskip\
\newblock
\APACrefYearMonthDay{2008}{}{}.
\newblock
{\BBOQ}\APACrefatitle {{Production of cosmogenic $^{7}$Be isotope in the
  atmosphere: Full 3-D modeling}} {{Production of cosmogenic $^{7}$Be isotope
  in the atmosphere: Full 3-D modeling}}.{\BBCQ}
\newblock
\APACjournalVolNumPages{J. Geophys. Res.}{113}{D12}{D12107}.
\newblock
\begin{APACrefDOI} \doi{10.1029/2007JD009725} \end{APACrefDOI}
\PrintBackRefs{\CurrentBib}

\bibitem [\protect \citeauthoryear {%
{Usoskin}%
\ \protect \BOthers {.}}{%
{Usoskin}%
\ \protect \BOthers {.}}{%
{\protect \APACyear {2015}}%
}]{%
usoskin_PAMELA_15}
\APACinsertmetastar {%
usoskin_PAMELA_15}%
\begin{APACrefauthors}%
{Usoskin}, I\BPBI G.%
, {Kovaltsov}, G\BPBI A.%
, {Adriani}, O.%
, {Barbarino}, G\BPBI C.%
, {Bazilevskaya}, G\BPBI A.%
, {Bellotti}, R.%
\BDBL {}{Zverev}, V\BPBI G.%
\end{APACrefauthors}%
\unskip\
\newblock
\APACrefYearMonthDay{2015}{}{}.
\newblock
{\BBOQ}\APACrefatitle {{Force-field parameterization of the galactic cosmic ray
  spectrum: Validation for Forbush decreases}} {{Force-field parameterization
  of the galactic cosmic ray spectrum: Validation for Forbush
  decreases}}.{\BBCQ}
\newblock
\APACjournalVolNumPages{Adv. Space Res.}{55}{}{2940-2945}.
\newblock
\begin{APACrefDOI} \doi{10.1016/j.asr.2015.03.009} \end{APACrefDOI}
\PrintBackRefs{\CurrentBib}

\bibitem [\protect \citeauthoryear {%
{Usoskin}%
, {Mironova}%
, {Korte}%
\BCBL {}\ \BBA {} {Kovaltsov}%
}{%
{Usoskin}%
\ \protect \BOthers {.}}{%
{\protect \APACyear {2010}}%
}]{%
Usoskin_JASTP2010}
\APACinsertmetastar {%
Usoskin_JASTP2010}%
\begin{APACrefauthors}%
{Usoskin}, I\BPBI G.%
, {Mironova}, I\BPBI A.%
, {Korte}, M.%
\BCBL {}\ \BBA {} {Kovaltsov}, G\BPBI A.%
\end{APACrefauthors}%
\unskip\
\newblock
\APACrefYearMonthDay{2010}{}{}.
\newblock
{\BBOQ}\APACrefatitle {{Regional millennial trend in the cosmic ray induced
  ionization of the troposphere}} {{Regional millennial trend in the cosmic ray
  induced ionization of the troposphere}}.{\BBCQ}
\newblock
\APACjournalVolNumPages{J. Atmos. Solar-Terrestr. Phys.}{72}{}{19-25}.
\newblock
\begin{APACrefDOI} \doi{10.1016/j.jastp.2009.10.003} \end{APACrefDOI}
\PrintBackRefs{\CurrentBib}

\bibitem [\protect \citeauthoryear {%
{Vos}%
\ \BBA {} {Potgieter}%
}{%
{Vos}%
\ \BBA {} {Potgieter}%
}{%
{\protect \APACyear {2015}}%
}]{%
Vos_ApJ2015}
\APACinsertmetastar {%
Vos_ApJ2015}%
\begin{APACrefauthors}%
{Vos}, E\BPBI E.%
\BCBT {}\ \BBA {} {Potgieter}, M\BPBI S.%
\end{APACrefauthors}%
\unskip\
\newblock
\APACrefYearMonthDay{2015}{}{}.
\newblock
{\BBOQ}\APACrefatitle {{New Modeling of Galactic Proton Modulation during the
  Minimum of Solar Cycle 23/24}} {{New Modeling of Galactic Proton Modulation
  during the Minimum of Solar Cycle 23/24}}.{\BBCQ}
\newblock
\APACjournalVolNumPages{Astrophys. J.}{815}{}{119}.
\newblock
\begin{APACrefDOI} \doi{10.1088/0004-637X/815/2/119} \end{APACrefDOI}
\PrintBackRefs{\CurrentBib}

\bibitem [\protect \citeauthoryear {%
{Webber}%
, {Higbie}%
\BCBL {}\ \BBA {} {McCracken}%
}{%
{Webber}%
\ \protect \BOthers {.}}{%
{\protect \APACyear {2007}}%
}]{%
Webber_JGR2007}
\APACinsertmetastar {%
Webber_JGR2007}%
\begin{APACrefauthors}%
{Webber}, W\BPBI R.%
, {Higbie}, P\BPBI R.%
\BCBL {}\ \BBA {} {McCracken}, K\BPBI G.%
\end{APACrefauthors}%
\unskip\
\newblock
\APACrefYearMonthDay{2007}{}{}.
\newblock
{\BBOQ}\APACrefatitle {{Production of the cosmogenic isotopes $^{3}$H,
  $^{7}$Be, $^{10}$Be, and $^{36}$Cl in the Earth's atmosphere by solar and
  galactic cosmic rays}} {{Production of the cosmogenic isotopes $^{3}$H,
  $^{7}$Be, $^{10}$Be, and $^{36}$Cl in the Earth's atmosphere by solar and
  galactic cosmic rays}}.{\BBCQ}
\newblock
\APACjournalVolNumPages{Journal of Geophysical Research (Space
  Physics)}{112}{A11}{A10106}.
\newblock
\begin{APACrefDOI} \doi{10.1029/2007JA012499} \end{APACrefDOI}
\PrintBackRefs{\CurrentBib}

\bibitem [\protect \citeauthoryear {%
{Wilcox}%
, {Hoskins}%
\BCBL {}\ \BBA {} {Shine}%
}{%
{Wilcox}%
\ \protect \BOthers {.}}{%
{\protect \APACyear {2012}}%
}]{%
Wilcox_QJRMS2012}
\APACinsertmetastar {%
Wilcox_QJRMS2012}%
\begin{APACrefauthors}%
{Wilcox}, L\BPBI J.%
, {Hoskins}, B\BPBI J.%
\BCBL {}\ \BBA {} {Shine}, K\BPBI P.%
\end{APACrefauthors}%
\unskip\
\newblock
\APACrefYearMonthDay{2012}{}{}.
\newblock
{\BBOQ}\APACrefatitle {{A global blended tropopause based on ERA data. Part I:
  Climatology}} {{A global blended tropopause based on ERA data. Part I:
  Climatology}}.{\BBCQ}
\newblock
\APACjournalVolNumPages{Q. J. R. Meteorol. Soc.}{138}{664}{561-575}.
\newblock
\begin{APACrefDOI} \doi{10.1002/qj.951} \end{APACrefDOI}
\PrintBackRefs{\CurrentBib}

\end{thebibliography}

\end{document}